\documentclass[twocolumn,traditabstract]{aa}

\usepackage{graphicx}
\usepackage{natbib}
\usepackage{epsfig}
\usepackage{lscape}
\usepackage{amsmath}
\usepackage{txfonts}
\usepackage[english]{babel}
\usepackage[titletoc,title]{appendix}
\usepackage{txfonts}
\usepackage{times}
\usepackage{enumitem}
\makeatletter \def\@biblabel#1{} \makeatother

\def\msun{{\rm M_{\odot}}}

\begin{document}

\title{Physical properties of AGN host galaxies as a probe of supermassive black hole feeding mechanisms}

  \author{M. Gatti, A. Lamastra, N. Menci, A. Bongiorno, F. Fiore}
   \offprints{}
   \institute{INAF - Osservatorio Astronomico di Roma, via di Frascati 33, 00040 Monte Porzio Catone, Italy.}
   \date{Received ; Accepted }
   \abstract{ Using an advanced semi-analytical model (SAM) for galaxy formation, we investigated the statistical effects of assuming two different mechanisms for triggering AGN activity on the properties of AGN host galaxies. We considered a first accretion mode where AGN activity is triggered by disk instabilities (DI) in isolated galaxies, and a second feeding  mode where galaxy mergers and fly-by events (interactions, IT) are responsible for producing a sudden destabilization of large quantities of gas, causing the mass inflow onto the central supermassive black hole (SMBH). The effects of including IT and DI modes in our SAM were studied and compared with observations separately to single out the regimes in which they might be responsible for triggering AGN activity. We obtained the following results: i) the predictions of our model concerning the stellar mass functions of AGN hosts point out that both DI and IT modes are able to account for the observed abundance of AGN host galaxies  with $M_* \lesssim 10^{11} \msun$; for more massive hosts, the DI scenario predicts a much lower space density than the IT model in every redshift bin, lying below the observational estimates for redshift $z>0.8$. ii) The analysis of the colour-magnitude diagram (CMD) of AGN hosts for redshift $z < 1.5 $ can provide a good observational test to effectively distinguish between DI and IT mode, since DIs are expected to yield AGN host galaxy colours skewed towards bluer colours, while in the IT scenario the majority of hosts are expected to reside in the red sequence. iii) While both IT and DI scenarios can account for AGN triggered in main sequence or starburst galaxies, DIs fail in triggering AGN activity in passive galaxies. The lack of DI AGN in passive hosts is rather insensitive to changes in the model describing the DI mass inflow, and it is mainly caused by the criterion for the onset of disk instabilities included in our SAM. iv) The two modes are characterized by a different duration of the AGN phase, with DIs lasting even on time scales
of $\sim $ Gyr, much longer than in the IT scenario, where the galaxy interaction sets the duration of the burst phase around $\sim 10^7 - 10^8$ yr. v) The scatter of the star formation rate $SFR-L_{bol}$ relation could represent another crucial diagnostics to distinguish between the two AGN triggering modes, since the DI scenario predicts an appreciably lower scatter (especially at low-to-intermediate AGN luminosities) of the relation than the IT scenario. vi) Disk instabilities are unable to account for the observed fraction of AGN in groups for $z \lesssim 1$ and clusters for $ z \lesssim 0.7$, while the IT scenario matches observational data well.
}
\authorrunning{M. Gatti et al.}
\titlerunning{Physical properties of AGN host galaxies as a probe of supermassive black hole feeding mechanisms}
   \maketitle

\section{Introduction}
Since they were first discovered, many steps have been made towards understanding the physics of active galactic nuclei (AGN). At present,  we know that the masses of supermassive black holes
(SMBH)  in the local Universe scale with the structural parameters of their host bulges (e.g. Kormendy \& Richstone 1995, Magorrian et al. 1998, Gebhardt et al. 2000, Ferrarese \& Merritt 2000, Marconi \& Hunt 2003, H\"aring \& Rix 2004, McConnell \& Ma 2013; for a recent review see Kormendy \& Ho 2013). We realized
that the growth of black holes (BHs) is mostly due to accretion of matter during AGN phases (Soltan 1982).
We observe a luminosity-dependent evolution of the AGN population (Fiore 2003, La Franca 2005, Ueda 2003, 2014), with low-luminosity AGN reaching the peak in their space density  at lower redshift than higher luminosity AGN,  which  is similar to the downsizing behaviour of star-forming galaxies (Cowie et al. 1996). 
Given this framerwork, the next step is to understand the triggering mechanisms responsible for BH accretion and  their relation with the cosmological evolution of galaxies.

Galaxy major mergers are widely believed to be the main triggering mechanism for bright quasi-stellar objects (QSOs, that is, AGN with bolometric luminosities L$_{bol}\gtrsim$10$^{46}$ erg s$^{-1}$) on the basis of both observational evidence (Disney et al. 1995; Bahcall et al. 1997; Kirhakos et al. 1999; Hutchings 1987; Yates et al. 1989; Villar-Martn 2010; 2012; Treister et al. 2012) and theoretical arguments (Hernquist 1989; Barnes \& Hernquist
1991, 1996; Mihos \& Hernquist 1994, 1996; Di Matteo et al.2005; Cox et al. 2008). However, a number of recent studies on the star-forming properties (Lutz et al. 2010; Shao et al. 2010; Rosario et al. 2012, Mullaney et al.  2012, Santini et al. 2012) and on the morphology of AGN hosts (Salucci et al. 1999; Grogin et al. 2005; Pierce et al. 2007; Georgakakis et al. 2009; Cisternas et al. 2010; Villforth et al. 2014;Rosario et al. 2013a; Silverman et al. 2011; Kocevski et al. 2012) indicates that more moderate levels of nuclear activity  (Seyfert-like galaxies with L$_{bol}\lesssim10^{46}$erg s$^{-1}$) can be associated with galaxies undergoing secular evolution rather than major mergers, suggesting that also internal processes such as disk instabilities in isolated galaxies might play a relevant role in triggering AGN activity. Interestingly, it seems that the different triggering mechanisms are closely linked to different physical processes driving the star formation activity in the host galaxies, with starburst galaxies being induced by major mergers, while internal processes are responsible for driving typical star-forming galaxies on the galaxy main sequence (Noeske et al. 2007; Elbaz et al. 2007; Daddi et al. 2007, 2009; Santini et al. 2009).

A detailed statistical study of the role of different triggering mechanisms and their relation with the properties of the host galaxies can be performed using cosmological models of galaxy formation. 
Analytical descriptions of both galaxy interactions (i.e. merging events and tidal interactions, hereafter IT) and disk instability (hereafter DI) as physical triggers for AGN activity have been implemented in several semi-analytic models (SAM,  Croton 2006; Bower et al. 2006; Hirschman et al. 2012, Fanidakis et al. 2012) or in N-body simulations as sub-grid physics (Springel et al. 2005;  Di Matteo et al. 2005; Li et al. 2007; Hopkins et al. 2010). 

The results of these studies depend on the particular analytical prescriptions assumed for the AGN feeding modes. For the IT scenario, several authors have worked out analytical or phenomenological descriptions of the mass inflow induced by galaxy interactions (see, e.g., Makino \& Hut 1997, Cavaliere \& Vittorini 2000), with these models having been calibrated and tested against aimed hydrodynamical merger simulations (Robertson et al. 2006a,b; Cox et al. 2006; Hopkins et al. 2007b). A different situation holds for the DI scenario: so far, the  inflows driven by disk instability have been implemented into SAMs only through a simplified \textit{ansatz} concerning the fate of disk gas and stars when the disk becomes unstable  (e.g.  Hirshman \& Somerville 2012; Fanidakis et al. 2012) because we still lack an accurate analytic theory predicting gas inflow rates in unstable stellar-plus-gas galaxy disks. Recently, however, a step forward in describing gas inflow due to disk instabilities into SAMs has been presented by Menci et al. (2014, hereafter M14). In M14 we included the Hopkins \& Quataert (2011, HQ11 hereafter) model for disk instability in an updated version of the Rome SAM of galaxy formation (Menci et al., 2006 2008). The HQ11 model is an analytic model that attempts to account for the physics of angular momentum transport and gas inflow triggered by disk instabilities in realistic stellar-plus-gas systems, following the inflow from galactic scales ($ \thickapprox $1 kpc) inward to nuclear scales ($\lesssim$ 10 pc) and relating it to nuclear star formation activity. This model has been tested by the authors against aimed hydrodynamical simulations, so that at present it constitutes a solid baseline to describe the BH accretion due to disk instabilities.

In M14  we derived the luminosity and redshift intervals where the two different feeding modes might be effective by comparing the AGN luminosity function up to $z=$6 with that predicted by the SAM assuming 
only DI as a trigger for AGN activity or assuming a pure interaction-driven feeding mode. We found that even when we maximized the BH accretion rate due to DI, DI cannot provide the observed abundance of high-luminosity QSO with  $M_{1450} < -26$ and of high-redshift $z \gtrsim 4.5 $ AGN with $M_{1450} < -24$. On the other hand, DI constitues a viable candidate mechanism to fuel AGN with $M_{1450} \gtrsim -26$  up to $z \sim 4.5$, effectively competing with the IT scenario in driving the cosmological evolution of the AGN population. Thus, identifying key observables able to distinguish which AGN fuelling mechanism is dominant in these low-to-intermediate luminosity and redshift ranges is relevant  for understanding AGN triggering mechanisms. In M14 we have shown that while the comparison with the local black hole mass function has proven to be inconclusive
(both IT and DI scenarios show very similar distributions that
are compatible with the present observational constraints), interesting discriminants could be provided by the $M_{BH} - M_{*}$ relation and by the Eddington ratio ($\lambda$)  distribution. As for the former, DI and IT show a very different scatter of the relation; although even in this case the predictions of both modes are still compatible with the latest observational constraints and numerical results (e.g, Kormendy \& Ho 2013, Sijacki et al. 2014), this difference in the scatter provides useful insights into some interesting features of the two scenarios. Indeed, the tightness that characterizes the $M_{BH} - M_{*}$ relation obtained with the DI mode has been related to the fact that in this scenario BH growth and star formation are tightly related, and phases where either $\dot{M}_{BH}$ or $\dot{M}_*$ largely dominate are absent. In contrast, the IT mode is characterized by a larger scatter in the $M_{BH}-M_*$ relation (due to the variance of the merging histories),  and phases where only $\dot{M}_{BH}$ dominates are more likely to be found.  This means that we expect a tight relation between AGN luminosity and SFR in the DI scenario and a larger scatter for the same relation for the IT mode. For the same reason, we expect that it would be very unlikely to find DI-driven AGN in passive, red hosts. 
We note here that similarly to the $M_{BH} - M_{*}$ relation, the $M_{BH}-\sigma$ relation (and its scatter) could also
represent a good diagnostic; however, since SAMs are not N-body simulations and the dispersion velocity is not a direct output of our model, we have chosen not to compare with this observable, since our results would unavoidably depend on the analytical prescriptions assumed to compute the velocity dispersion from some known galaxy parameter.
The insights provided by the Eddington ratio distributions can also be further investigated: since the DI model predicts a relatively narrow $\lambda$ distribution, we expect that more luminous AGN were found in more massive galaxies and less luminous AGN in lower mass galaxies.  Since DIs are unable to reproduce the bright end of the luminosity function, we expect that they also  fail  in triggering AGN activity in the most massive galaxies. The two models also have a different morphological dependence: DI can only be triggered in gas-rich galaxies with a marked disk component, while IT are less affected by morphology requirements. Since galaxies with different morphologies reside in different environments (for instance, the most massive clusters are mainly populated by ellipticals), an environmental dependence might be found.

The picture depicted above strongly suggests that a systematic study of AGN host galaxy properties might help to further constrain the physical regimes in which DI and IT mode might operate.
In this paper we test the DI and IT modes for AGN feeding by comparing the model predictions with the observed  colour distribution, star-forming properties, and environmental dependence of AGN hosts. 
The paper is organized as follows: in Sect. 2 we recall the basic properties of the Rome SAM for galaxy formation, focusing in particular on the DI and IT mode for AGN feeding included in our SAM. In Sect. 3 we present and discuss our results concerning several AGN host galaxy properties, obtained by comparing the predictions of our model with the latest available observational data; Sect. 4 summarizes our most relevant results.

\section{Semi-analytical model}
We base our analysis on the semi-analytical model (SAM) described in Menci et al. (2006, 2008; see M14 for the latest updates), which connects within a cosmological framework the merging tree of DM halos with the processes involving their baryonic content (such as gas cooling, star formation, supernova
feedback, and chemical enrichment). The backbone of the model is constituted by the merging trees of dark matter halos, generated through a Monte Carlo procedure on the basis of the merging rates of the DM halos provided by the extended
Press \& Schechter formalism (see Bond et al. 1991; Lacey
\& Cole 1993). The clumps included in larger DM halos may
survive as satellites, or merge to form larger galaxies as a
result of binary
aggregations, or coalesce into the central dominant galaxy as
a result of dynamical friction; these processes take place over time scales that
grow longer over cosmic time, so the number of satellite galaxies
increases as the DM host halos grow from groups to clusters (see
Menci et al. 2006).

The baryonic processes taking place in each dark matter halo are computed following the standard recipes commonly adopted in SAMs; the gas in the halo, initially set to have
a density given by the universal baryon fraction and to be at the virial temperature, cools as a result of atomic processes and settles into a
rotationally supported disk with mass $M_c$, disk radius $R_d$ and disk circular velocity $V_d$ computed as in Mo, Mao \& White (1998). The cooled gas $M_c$ is gradually converted into stars through the processes described in detail in Sect. 2.2. Part of the energy released
by supernovae (SNae) following star formation is fed back onto the galactic
gas, thus returning part of the cooled gas to the hot phase.

The luminosity - in different bands - produced by the stellar populations
of the galaxies are computed by convolving the star formation histories of the galaxy progenitors with a synthetic spectral energy
distribution, which we take from Bruzual \& Charlot (2003) assuming a Salpeter initial mass function (IMF).

The model also includes a recent description of the tidal stripping of
stellar material from satellite galaxies. The treatment adopted here is the same as was introduced by Henriques \& Thomas (2010) in the Munich
SAM, to which we refer for further details.

\subsection{AGN triggering}
In our SAM two different modes for the BH accretion rate are included: i) an interaction-triggered (IT) mode, where the triggering of the AGN activity is {\it external} and provided by galaxy interactions (including mergers and fly-by events); ii) a mode 
where accretion occurs as a result of disk instabilities (DI), where the trigger is {\it internal} and provided by the break of the axial symmetry in the distribution of the galactic cold gas. In both cases, nuclear star formation is associated with the accretion flow triggering the AGN. 

The implementation of the IT mode has been described and extensively tested in previous papers (Menci et al. 2006, 2008, Lamastra et al. 2013b), while the DI mode is discussed in depth in M14, which we refer to for more details. Here, we give a basic descriptions for both IT and DI models.

As for the interaction scenario, galaxy interactions in a galactic halo with given circular velocity $v_c$ inside a host halo with circular velocity V  occur at a rate
\begin{equation}\label{int}
\tau_r^{-1}=n_T(V)\,\Sigma (r_t,v_c,V)\,V_{rel} (V)
,\end{equation}
where $n_T=3 N_T/4\pi R_{vir}^3$ is the number density of galaxies in the host halo, $V_{rel}$ the relative velocity between galaxies, and $\Sigma$ the cross section for such encounters, which is given by Saslaw (1985) in terms of the tidal radius $r_t$ associated with a galaxy with given circular velocity $v_c$ (Menci et al. 2004). When two galaxies merge, the surplus of cold gas is added to the galaxy disk of the galaxy resulting from the merging event. We have recently introduced a more detailed treatment of the transfer of stellar mass to the bulge during major mergers, following Hopkins et al. (2009): in particular, we assume that in mergers with $M_{*1}/M_{*2} \gtrsim 0.2$ only a fraction $1 - f_{gas}$ of the disk mass is transferred to the bulge. The fraction $f$ of cold gas destabilized by any kind of merging or interaction has been calculated by  Cavaliere \& Vittorini (2000)  in terms of the  variation $\Delta j$ of the specific angular momentum $j\approx
GM/V_d$ of the gas as (Menci et al. 2004)
\begin{equation}\label{fdest}
f\approx \frac{1}{2}\,
\Big|{\Delta j\over j}\Big|=
\frac{1}{2}\Big\langle {M'\over M}\,{R_d\over b}\,{V_d\over V_{rel}}\Big\rangle\, 
,\end{equation}
where $b$ is the impact parameter, evaluated as the greater of the radius $R_d$ and the average distance of the galaxies in the halo, $M'$ is the mass of the  partner galaxy in the
interaction, and the average runs over the probability of finding such a galaxy
in the same halo where the galaxy with mass $M$ is located.  The pre-factor accounts for the probability that half of the
inflow rather than of the outflow is related to the sign of $\Delta j$. We assume that in each interactions a quarter of the destabilized  fraction $f$ feeds the central BH, while the remaining part feeds the circumnuclear starbursts  (Sanders \& Mirabel 1996). Thus, the BH accretion rate is given by 
\begin{equation}
\label{macc_ID}
 {dM_{BH}\over dt}={1\over 4}{f\,M_c\over \tau_b}  
,\end{equation}
where the time scale $\tau_{b}=R_d/V_d$  is assumed to be the crossing time of the destabilized galactic disk. The duration of an accretion episode, that is, the time scale for the QSO or AGN to shine, is equal to the crossing time of the destabilized galactic disk ($\tau_b$).

For the DI scenario, disk instability arises in galaxies whose disk mass exceeds a given critical value
\begin{equation}
\label{efstathi}
M_{crit} =  {v_{max}^2 R_{d}\over G \epsilon}
,\end{equation} 
where $v_{max}$ is the maximum circular velocity, $R_d$ the scale length of the disk, and $\epsilon$ a parameter in range $\epsilon\sim 0.5 - 0.75$;  we adopt the value $\epsilon=0.75$ for the latter
(we note that the same value has also been adopted by other SAMs, e.g. Hirschmann et al. 2012).  The above critical mass is provided by Efstathiou et al. (1982) on the basis of N-body simulations. For each galaxy, for each time step of the simulation, we compute the critical mass following Eq. \ref{efstathi}; if the criterion is satisfied, the perturbation settles down, and we proceed in computing the mass inflow generated by disk instabilities according to the HQ11 model. The final inflow onto the central BH is equal to
\begin{multline}
\label{hopkins}
\frac{dM_{BH}}{dt}  \approx 
{\alpha \,
f_d^{4/3}\over 1+2.5\,f_d^{-4/3}(1 + f_0/f_{gas}) } \times \\
\left( \frac{M_{BH}}{10^8 M_{\bigodot}}\right)^{1/6}
\left( \frac{M_d(R_0)}{10^9 M_{\bigodot}}\right)
\left( \frac{R_0}{100 pc }\right) ^{-3/2}
M_{\bigodot} yr^{-1}
,\end{multline}
where 
\begin{equation}
f_0 \approx 0.2 f_d^2 \left[ \frac{M_d(R_0)}{10^9 M_{\bigodot}}\right]^{-1/3}
{\hspace{2cm}}  f_{gas} \equiv {M_{gas} (R_0)\over M_d(R_0)}
.\end{equation} 
Here $M_{BH}$ is the central black hole mass,  $f_d$ is the total disk mass fraction, $M_d$ and $M_{gas}$ the disk and the gas mass calculated in $R_0$ (we take $R_0 = 100$ pc).

The constant $\alpha$ parametrizes several uncertainties related to some of the basic assumptions of the HQ11 model; its value is not completely freely tunable, but the limits imposed by the HQ11 model set the range of acceptable values to be approximately $\alpha$ = 0.1 $\sim$ 5 (see M14 for more details). In spite of this range of validity, in M14 we have considered a highest value of $\alpha$ = 10 to assess whether the DI scenario might
be able to reproduce the bright end of the AGN luminosity function or not, even maximizing the accretion rate onto the central BH; in this paper, however, we can also explore the effects of considering lower values for the parameter $\alpha$. We caution that lowering the normalization of the accretion rate strongly affects the predicted AGN luminosity function, reducing even more the luminosity and redshift range where DIs might be able to trigger AGN activity: for instance, taking $\alpha=2$, the DI scenario is able to reproduce the observed AGN luminosity function only for $z \lesssim 4$ and for luminosity $M_{1450} \gtrsim -23$. Moreover, a lower normalization also affects the AGN duty cycle, which is now lengthened (see Sect. 3.3), resulting in an overproduction of very low-luminosity AGN. For these reasons, we have chosen not to further lower the value of the parameter $\alpha$ in our analysis, considering the case of $\alpha=2$ as a lower limit for the normalization of the BH accretion rate described by Eq. \ref{hopkins}. In the following we show for the DI scenario both the predictions for the case with maximized normalization $\alpha=10$ (the $"$fiducial$"$ case, also considered in M14) and the case with $\alpha=2$, so as to span all the reasonable values of the normalization of the mass inflow predicted by the HQ11 model.

To compare our predictions with observations, we converted the BH mass inflows into AGN bolometric luminosity using the following equation:
\begin{equation}\label{Lagn}
L_{AGN}=\eta\,c^2\,{d M_{BH}\over dt}
,\end{equation}
where $dM_{BH}/dt$ is taken from Eq. \ref{macc_ID} for the IT mode and from Eq. \ref{hopkins} for the DI scenario. We adopted an energy-conversion efficiency $\eta= 0.1$ (see Yu \&
Tremaine 2002). The luminosities in the UV and in the X-ray bands
were computed from the above expression using the bolometric correction given in Marconi et al. (2004).

Our SAM also includes a detailed treatment of AGN feedback; the feedback model included in our SAM is the blast wave model for AGN feedback developed by Cavaliere et al. (2002) and Lapi et al. (2005) (Menci et al., 2008). According to the model, the supersonic outflows generated by AGN during their luminous phase compress the gas into a blast wave that moves outwards, eventually
expelling a certain amount of gas from the galaxy. Quantitatively, after the end of AGN activity, the gas content of the galaxy is depleted by a factor $ 1 - \Delta m/m \thickapprox 1 - \Delta E /2E$ , where $\Delta E $ is the energy injected by the AGN into the galactic gas during each accretion episodes and $E$ is the gas-binding energy of the interstellar medium (ISM). 
 
\subsection{Star formation rate}
In our SAM, different channels of star formation
may convert part of the gas content of a galaxy into stars.\\
\begin{enumerate}[label=\roman*)]
\item First of all, all galaxies form stars with a rate equal to 
\begin{equation}
\label{quie}
SFR_q =\frac{M_c}{q t_d}
,\end{equation}
where $m_c$ is the mass of cold gas contained in the galaxy, $t_d = r_d/v_d$ is the dynamical time of the disk, and $q$ an adjustable parameter chosen to match the Kennicutt (1998) relation. In the following, we refer to this mode of star formation as quiescent.
\item Secondly, galaxy interactions are responsible for triggering an additional SF burst, equal to 

\begin{equation}
\label{burst1}
SFR_{b,I}=p \, {f\,M_c\over \tau_b}  
,\end{equation}
where $f$, $M_c$ and $\tau_b$ are the quantities defined in Eq. \ref{macc_ID}. In the IT scenario, the parameter $p$ is taken to be $p=3/4$; in fact, while $ \sim 1/4$ of the fraction of gas destabilized during the galaxy interaction is accreted onto the central BH, the remaining $\sim 3/4$ is assumed to feed the circumnuclear starburst. Because galaxy interactions \textit{do not} trigger any AGN activity, all the destabilized gas in the DI\ scenario is assumed to feed the circumnuclear starburst, and therefore we take $p=1$.
\item In addition to the two SF modes mentioned above, because the HQ11 DI model assumes an equilibrium between the mass inflow and star formation, a star formation activity $SFR_{b,D}$ is always connected to the BH acccretion rate of Eq. \ref{hopkins} in the DI scenario. The value of the $SFR_{b,D}$ is computed in detail in M14 and reads
\begin{equation}
\label{burst2}
SFR_{b,D}=100 \, \dot{M}_{BH}
.\end{equation} 
The stars produced during the nuclear starbursts are added to the bulge. We note that the value $SFR_{b,D}=100 \dot{M}_{BH}$ represents a lower limit for the burst SFR within the assumptions of the HQ11 model: in fact, in M14 we have shown that this follows from considering only the contribution to the star formation rate from the inner region of the galaxy where the BH dominates the potential. Considering the contribution from the outer region
as well would result in a higher value of the $SFR_{b,D}$; however, since the exact value of the latter contribution is uncertain because of its strong dependence on the assumed disk surface density profile, we only consider the lower limit expressed by Eq. \ref{burst2} throughout this paper and discuss the effects of an increased $SFR_{b,D}$ when necessary.

\end{enumerate}

\section{Results}

Following M14, we stress that the aim of this work is to single out the effects of the IT and DI scenario on the properties of AGN hosts; hence, instead of trying to develop a best-fitting model by tuning the relative role of the two feeding modes, the effects of including IT and DI modes in our SAM were studied and compared with observations \textit{separately}. In the following sections, we proceed in testing the predictions of the two different feeding modes, focusing on the properties of AGN host galaxies.

\subsection{Stellar mass function of AGN host galaxies}

\begin{figure*}
\centering
\includegraphics[scale=0.6]{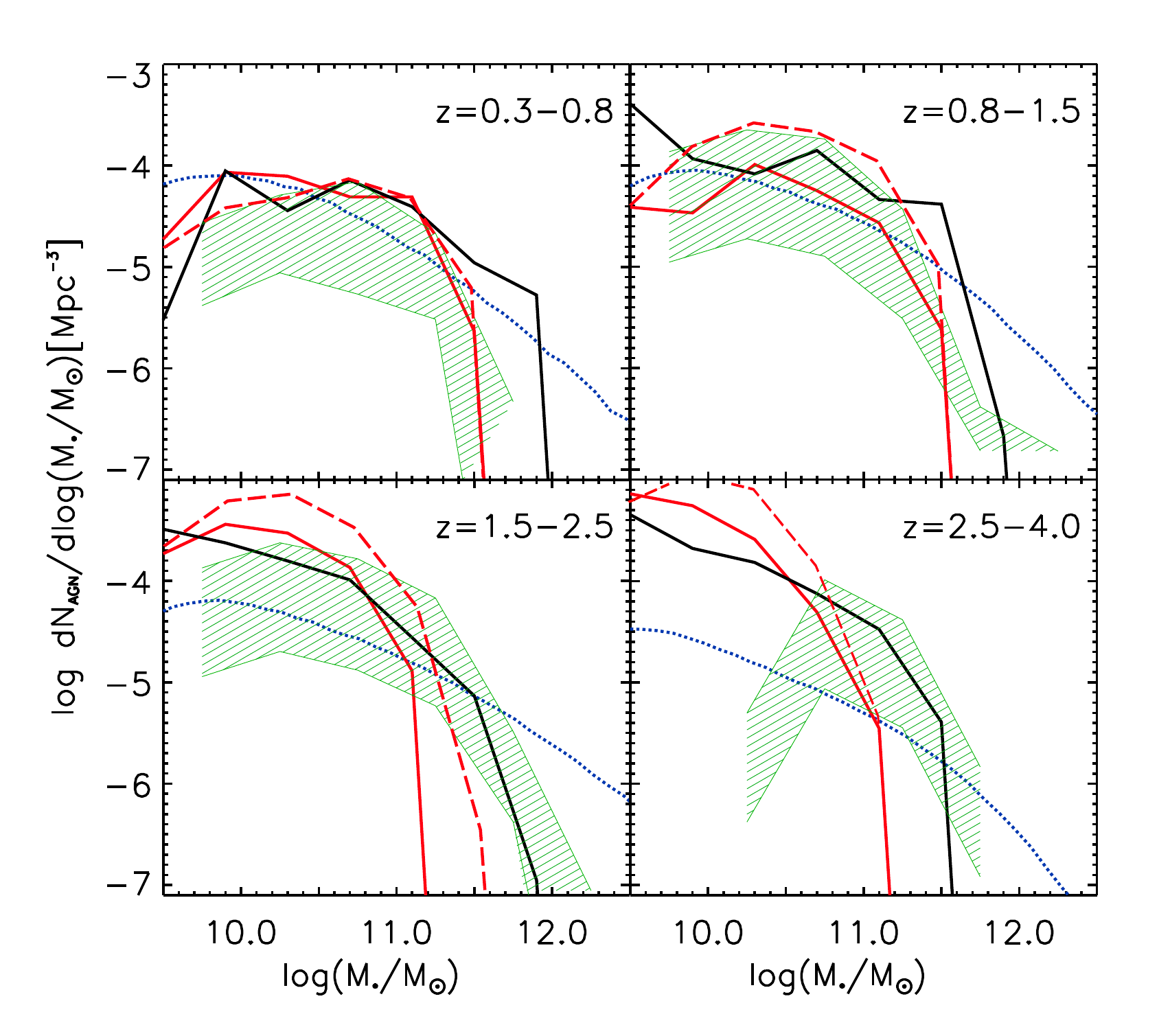}
\caption{Host stellar mass function for AGN with bolometric luminosity $L_{X} > 10^{43} $erg s$^{-1}$ for four different redshift bins. For the IT scenario, the black solid line represents the host stellar mass function predicted by the model; for the DI scenario, the red solid line represents the predictions for the fiducial case ($\alpha=10$), while the red dashed line represents the case with the normalization of the accretion rate decreased ($\alpha=2$). The blue dotted line represents the host stellar mass function estimate from Fiore et al. (2012) as computed by the authors at redshift 0.75, 1.25, 2, and 3.5. The shaded region represents the observational data from Bongiorno et al. (in prep.); the contours of the region are obtained by using different assumptions about the mass incompleteness of the sample. In particular, the upper limit is obtained by assuming an Eddington ratio $\lambda$ distribution for the sample described by a power law (as computed in Bongiorno et al. 2012), with a cut-off at $\lambda \sim 1 \%$, while the lower limit assumes the same distribution with a cut-off at $\lambda \sim 10 \%$.}
\label{massfunctionagn}
\end{figure*}


The first observable we compared our results with is the AGN host stellar mass function, which is shown in Fig. \ref{massfunctionagn} for AGN with luminosity in the 2-10 keV band $L_{X} > 10^{43} $erg s$^{-1}$ in four different redshift bins.

The predictions of our SAM are compared with data from Bongiorno et al. (in prep.), who consider a sample of X-ray selected AGN from the  XMM-COSMOS field (Hasinger et al., 2007, Brusa et al., 2010). The stellar masses were derived  by fitting the observed spectra with a two-component spectral energy distribution (SED)
fitting model, based on a combination of AGN spectra templates and host-galaxy models, as described in detail in  Bongiorno et al. (2012).
The AGN host mass function is obtained using the $V_{max}$ estimator (Schmidt 1968), which gives the space-density contribution of individual objects. The sample was corrected  for both AGN X-ray luminosity incompleteness and for mass incompleteness. The latter arises because given a certain $L_X$-cut/flux limit, an AGN with low Eddington ratio is only included in the AGN sample if its $M_{BH}$ is massive (and thus if it is hosted by a massive galaxy). To correct for this mass incompleteness, the authors assume an Eddington ratio $\lambda$ distribution described by a power law (as computed in Bongiorno et al. 2012), with a cut-off at low Eddington ratio ($\lambda \sim 1 \%$ or $\lambda \sim 10 \%$, see figure caption). We caution that these data are preliminary and may be subjected to changes (especially for the treatment of the mass incompleteness and the assumed Eddington ratio distribution).

We also compared our results with the empirical estimates of the host stellar mass function from Fiore et al. (2012). The authors first converted the X-ray AGN luminosity functions (taken from La Franca et al. 2005 for $z \lesssim 2$ and from their work for $z > 3$) into BH mass functions using Monte Carlo realizations and assuming a given Eddington ratio distribution. They used log-normal $\lambda$ distributions (Netzer 2009b, Trakhtenbrot et al. 2011, Shemmer et al. 2004, Netzer \& Trakhtenbrot 2007, Willott et al. 2010b), with the position of the peak shifting towards higher values of $\lambda$ with increasing redshift, ranging from $\lambda \sim 0.03$ at $z < 0.3$ to $\lambda \sim 0.22$ at $z \sim 3-4$. The BH mass functions were then converted into host stellar mass functions using a $M_{BH}$-$M_*$ relation. They assumed   $\Gamma_0$=log($M_{BH}$/$M_*$)= -2.8 at $z \sim$ 0 (H\"aring \& Rix 2004), with a redshift evolution described by $\Gamma= \Gamma_0 (1+z)^{0.5}$. Because a clear distinction between bulge and disk is still debated at high redshift (at high redshift, galaxy disks are much more compact and thicker than currently observed spirals), they assumed that the SMBH mass is only proportional to half of the total stellar mass of the galaxy, not to the total stellar content.

The comparison between the model predictions and the observations indicates that while for galaxies with $M_* \lesssim 10^{11} \msun$ both the DI model and IT model are able to  account for the observed abundance of AGN hosts, for more massive hosts the DI scenario predicts a much lower space density than the IT model in every redshift bin, lying below the observational estimates for redshift $z>0.8$. In the lowest redshift bin, however, DIs are still able to reproduce the observational estimates from Bongiorno et al., apparently providing a better match than the IT scenario.

The fact that DIs predict a lower space density of massive hosts than galaxy interactions can be explained as follows: in our model the most massive galaxies are formed in biased regions of the primordial density field that have already converted most of their gas into stars as a result of high-z interactions and merging events. The relatively low gas fraction and the high B/T ratio (due to the high rate of interactions experienced) in the most massive hosts suppress the AGN activity predicted by the DI mode: in fact, the mass inflow described by Eq. \ref{hopkins} is strongly reduced in case of low disk fractions $f_d$ or in case of low gas supply $f_{gas}$. Moreover, if galaxies are not disky enough, the Efstathiou criterion (Eq. \ref{efstathi}) prevents DI from triggering AGN activity. Although more gas is available in the highest redshift bin, the increasing number of interactions predicted by standard SAMs continuously disrupts forming disks, which again prevents DI from triggering AGN activity. 

However, the comparison with observational data is hampered by the large uncertainties that still affect the host stellar mass function estimates. For instance, the data from Bongiorno et al. are affected by the uncertainties related to the host galaxy stellar mass estimates and by the assumptions concerning the incompleteness function; particularly, the Eddington ratio distribution assumed to derive the mass incompleteness represents a critical problem. While our predictions generally agree well with the observational data when they are corrected assuming an Eddington ratio distribution with a $\lambda \sim 1 \%$  cut-off (upper limit of the shaded region), they constantly overpredict the abundance of AGN hosts for a $\lambda \sim 10 \%$ cut-off (lower limit of the shaded region). We also note that while the cut-off $\lambda=10 \%$ is very conservative and might be considered as a good lower limit for the AGN host mass function, the upper limit of the shaded region shown in Fig. \ref{massfunctionagn} obtained by considering a cut-off of $\lambda \sim 1 \%$ may be not as restrictive, as several authors have pointed out a large spread in the Eddington ratio distribution that also extends to values lower than $\lambda=0.01$ (e.g. Shankar 2013, Heckman 2014). Assuming a broader distribution in computing the mass incompleteness - at least for the most massive, passive hosts - would result in higher abundance of massive hosts, which might reduce the mismatch with IT predictions in the lowest redshift bin. The Eddington ratio distribution also critically affects the host mass function estimates from Fiore et al. (2012); the authors note that the normalization at $M_* = 10^{11} M\odot$ changes by a factor of 15-30 $\%$ and $\sim$ 100$ \%$ for a distribution peak and width differing by 30 $\%$ from the assumed normalizations. At high redshift, the uncertainties related to the X-ray AGN luminosity function or possible deviations in the SMBH mass-bulge mass relation can also affect the host mass function calculations, especially at low masses. 

Given this complex observational situation, more observational data are needed to effectively constrain the two feeding modes through the stellar mass function of AGN host galaxies, and both IT and DI scenarios can still be considered viable candidates for feeding moderately luminous AGN hosted in medium-sized galaxies. Hence, to clarify this question, we proceed in comparing the two modes with other host galaxy properties.

\subsection{Colour-magnitude diagram}

\begin{figure*}
    \centering
\includegraphics[scale=0.58]{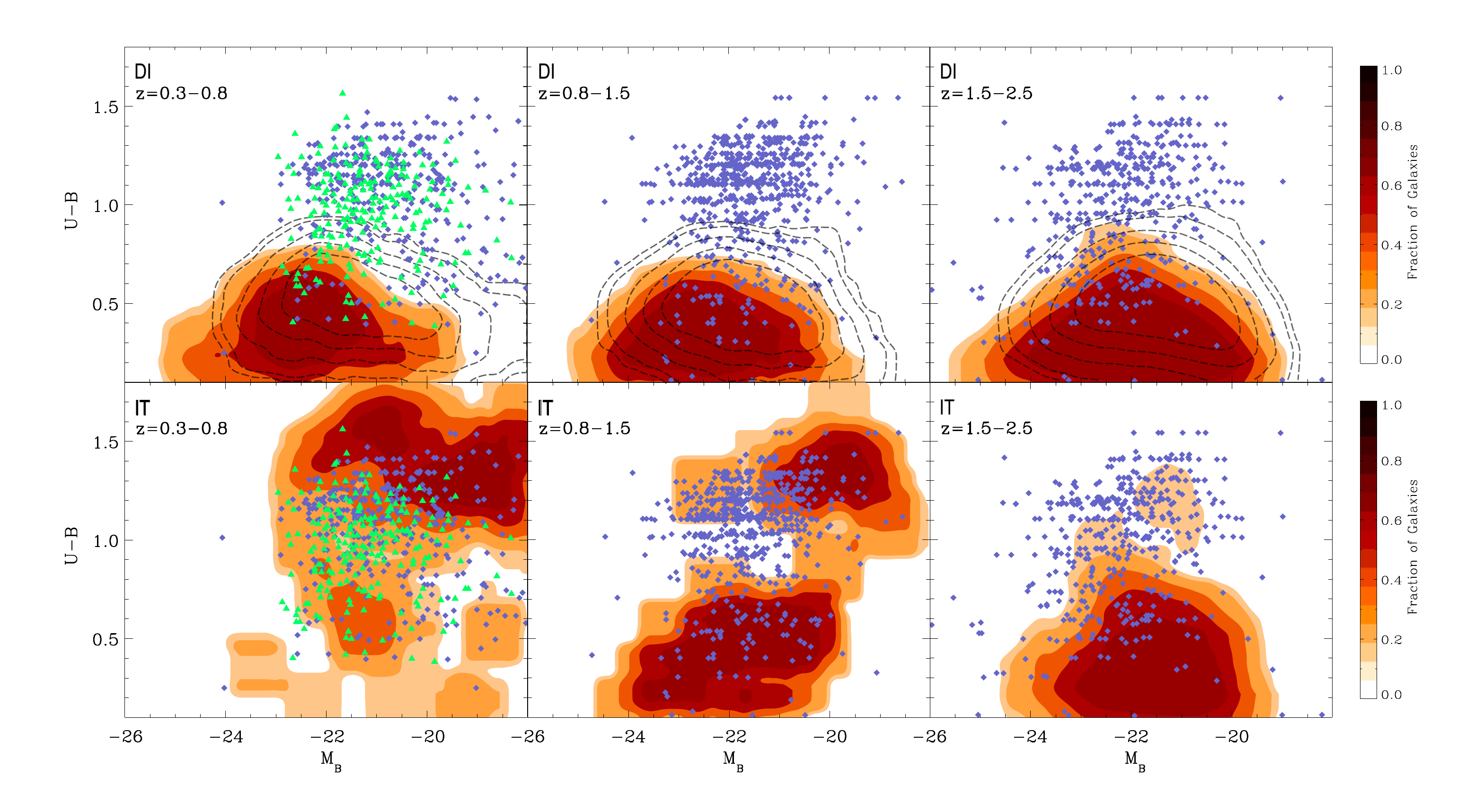}
\caption{Predicted CMD diagrams for galaxies hosting AGN with X-ray luminosity $L_X > 10^{42.2} \,$ erg s$^{-1}$ on three redshift bins are compared with the data from the COSMOS catalogue (Bongiorno et al. 2012, diamonds) and from Aird et al. 2012 (green triangles). Top panels refer to the DI model, with the predictions for the
fiducial case ($\alpha=10$) represented by the solid contour plot, while the grey dashed contour plot refers to the case $\alpha=2$. Bottom panels refer to the IT scenario.}
\label{CMD}
\end{figure*}

A useful tool that might allow distinguishing between the IT and DI scenario is provided by the colour-magnitude diagram (CMD).  The galaxy colour could reveal much about AGN host galaxy properties: the position on the CMD depends on the age of the stellar population, on its metallicity, and on extinction by dust. This in turn relates to the star-forming properties of the galaxies and to the galaxy morphology.

In the past years, several studies have shown that the colour function of galaxies in the local Universe is characterized by a bimodal distribution (Baldry et al. 2004): galaxies are inclined to reside in two well-defined regions of the CMD, the so-called red sequence and the blue cloud. The
red sequence mainly consists of early-type, massive, non-star-forming galaxies that have already built
the larger part of their mass at higher redshifts; on the other hand, the blue cloud
is characterized by less massive, late-type galaxies that are currently forming stars. In addition to these two populations of galaxies, an intermediate population lies in between in the so-called
green valley, which is generally interpreted as a region populated by galaxies that are quenching their star formation.

However, when the AGN hosts are considered, the situation is more complex, and many previous studies have often obtained contrasting conclusions with respect to AGN distribution in the CMD. 
Nandra et al. (2007) studied the positions of X-ray
AGN hosts in the CMD at z $ \sim $1 finding an overdensity of AGN in the luminous
galaxies close to and within the green valley. 
Coil et al. (2009) found analogous results for a larger sample of X-ray AGN, and the same trend has been noted also at lower redshifts (Hickox et al. 2009; Georgakakis \& Nandra 2011). 
Nevertheless, Silverman et al. (2009) and Xue et al. (2010) have shown the importance of using mass-matched samples in such analyses, indicating that stellar mass, not colour, may be the key parameter that drives the observed trends in the CMD. 
Xue et al. (2010) found that X-ray AGN are distributed over the full range of colours in stellar mass-matched samples with no specific clustering in the CMD in the redshift range $z$=1-4.
In contrast, an enhancement of X-ray selected AGN  in galaxies with blue or green  optical colours was found by Aird et al. (2012) even considering stellar-mass dependent selection effects. 
Cardamone et al. (2010), once having accounted for dust extinction effects in their sample, find that $\sim 25\%$ of red AGN and  $\sim 75\%$ of green AGN move to the blue cloud, leading to a bimodality distribution of AGN host galaxy colors. Finally, Bongiorno et al. (2012) noted that in addition to dust extinction, AGN emission might also affect AGN rest-frame optical colours: if not subtracted properly, the AGN component could be responsible for a significant blue contamination. They studied the distribution in the CMD of a sample of X-ray selected AGN from XMM-COSMOS field (Cappelluti et al 2009,  Hasinger et al. 2007) and purely optically selected AGN from the zCOSMOS bright survey (Lilly et al. 2009) and found a marginal  enhancement of the incidence of AGN in redder galaxies after
the colour-mass degeneracy in well-defined mass-matched samples was accounted for.  They argued that this result might emerge because of their ability to properly
account for AGN light contamination and dust extinction, compared to surveys with a more limited multiwavelength coverage.

In Fig. \ref{CMD} we show the predicted U-B rest-frame colours versus B-band absolute magnitude for galaxies hosting AGN with X-ray luminosity $L_X > 10^{42.2} \,$ erg s$^{-1}$ in three different redshift bins for DI and IT modes. The model predictions are compared with data from Bongiorno et al. (2012). In the lowest redshift bin, we also compare with data from Aird et al. (2012), who considered a sample of X-ray selected AGN from the XMM-COSMOS, XMM-LSS (Ueda et al. 2008) and XMM-ELAIS-S1 (Puccetti et al. 2006) fields with luminosities in the range $ 10^{42} < \,$ $L_X < 10^{44} \,$ erg s$^{-1}$ at $0.4 < z < 0.7$. We caution that the observational data from Aird et al. have been converted to $U-B$ and $M_B$ colours using a filter conversion provided by Blanton et al. (2007), which procedure might have introduced some additional uncertainties in the distribution of AGN hosts in the CMD. 

DI and IT modes produce very different distributions in the CMD, especially at low redshift. In the DI scenario, all redshift bins are characterized by AGN host galaxy colours skewed towards bluer colours; in contrast, the majority of host galaxies at low redshift in the IT\ scenario reside in the red sequence, gradually populating the blue cloud at higher redshifts. These behaviours can be explained by taking into account the different physical conditions that trigger AGN activity in the two modes. DI mode requires galaxies to be disky and gas rich (that is, highly star-forming galaxies characterized by young stellar populations); the Efstathiou criterion for DI triggering is only satisfied by galaxies with a marked disk component, while the net mass inflow onto the central BH (Eq. \ref{hopkins}) and AGN activity are strongly suppressed in case of lack of gas or with bulge-dominated galaxies. On the other hand, interaction-driven AGN triggering is less affected by morphology requirements and can also occur in old ellipticals or bulge-dominated galaxies where SF is quenched or has begun to diminish. 
Moreover, for the DI scenario the high ratio $SFR_{b,D}/\dot{M}_{BH}$ expressed by Eq. \ref{burst2} (which is substantially higher than the ratio $SFR_{b,I}/\dot{M}_{BH}$ for the IT scenario described by Eq. \ref{burst1}) always guarantees a significant burst of SF during the AGN phase, even for the less luminous AGN, which adds to the already high quiescent SF and might contribute to move AGN hosts even farther out in the blue region.
In this respect, we note that for $\alpha=2$ for the DI scenario, after decreasing the normalization of the mass inflow, we also
diminished  the value of the $SFR_{b,D}$ associated with AGN activity (again as a result of the proportionality between the BH accretion rate and burst SFR described by Eq. \ref{burst2}):  $\alpha=2$ produces slightly less luminous and less blue host colours than the fiducial case, which is due to a less prominent SF activity. In the highest redshift bin, the tendency of the IT mode to also occupy the blue cloud might be due to the higher fraction of gas available in the host galaxies at these redshifts, and thus due to a general enhancement of star formation.

In general, the IT scenario agrees quite well with observational data, while the DI mode is unable to populate the red sequence
sufficiently, showing a substantial mismatch. The mismatch is milder for $\alpha=2$, with DIs still being able to reproduce a substantial fraction of AGN, mainly residing in the blue cloud. We caution that some of the predicted features (such as the overprediction of red hosts in the lowest redshift bin for the IT mode or the extremely blue hosts in the DI scenario) might be alleviated by a better treatment of colour estimates and dust extinction by our model. We should also take into account possible non-optimal AGN-galaxy decompositions (or AGN over-subtractions) and a possible dependence on the specific choice of stellar population templates used to obtain the observational data. In Fig. \ref{CMD} we  note that we have compared our predictions with data that were not corrected for the effects of dust extinction, while Bongiorno et al. (2012) also provided the CMD for AGN hosts after they were de-reddened with the extinction derived from their SED fit. If we consider these latter data (lower panel of their Fig. 11), a substantial fraction of AGN hosts moves to the blue cloud; on the other hand, the effects of dust extinction on the predictions of our SAM are milder. The discrepancy between the effects of the dust extinction correction of our model and correction of the data is particular noticeable in the highest redshift bin and might explain why the match between the IT predictions and data becomes poorer in this redshift range. 
Similarly, the presence of the two well-defined clouds in the CMD at redshift z=0.8-1.5 for the IT scenario (representing a transient between the distributions at z = 0.3-0.8 and z=1.5-2.5), which are not shown by observational data, might be attributed to the previously mentioned uncertainties (concerning both the observational data and our predictions).

Even if the observational situation is still too uncertain to use the CMD to accurately assess the relative role of the two modes in reproducing AGN host colours, the predictions of our model strongly disfavour DIs as the dominant accretion mode for AGN hosted in red galaxies. Moreover, since the two scenarios show very different colour distributions (especially in the lowest redshift bin), future and unbiased studies of the AGN host galaxy colour might be useful to further constrain the regimes in which the two modes might operate.

The differences in the predicted CMD distributions suggest that differences in typical AGN host galaxy star formation rates may also be found: even though CMD cannot be trusted as a good tracer for star formation (e.g. Rosario et al. 2013b), the substantial lack of red AGN host galaxies in the case of DI mode might also
indicate a lack of passive host galaxies, which could be different for the IT mode. Thus, in the next section, we proceed in testing our AGN triggering modes against host galaxy star formation activity.

\subsection{$SFR-M_*$ relation}

\begin{figure*}
    \centering
\includegraphics[scale=0.6]{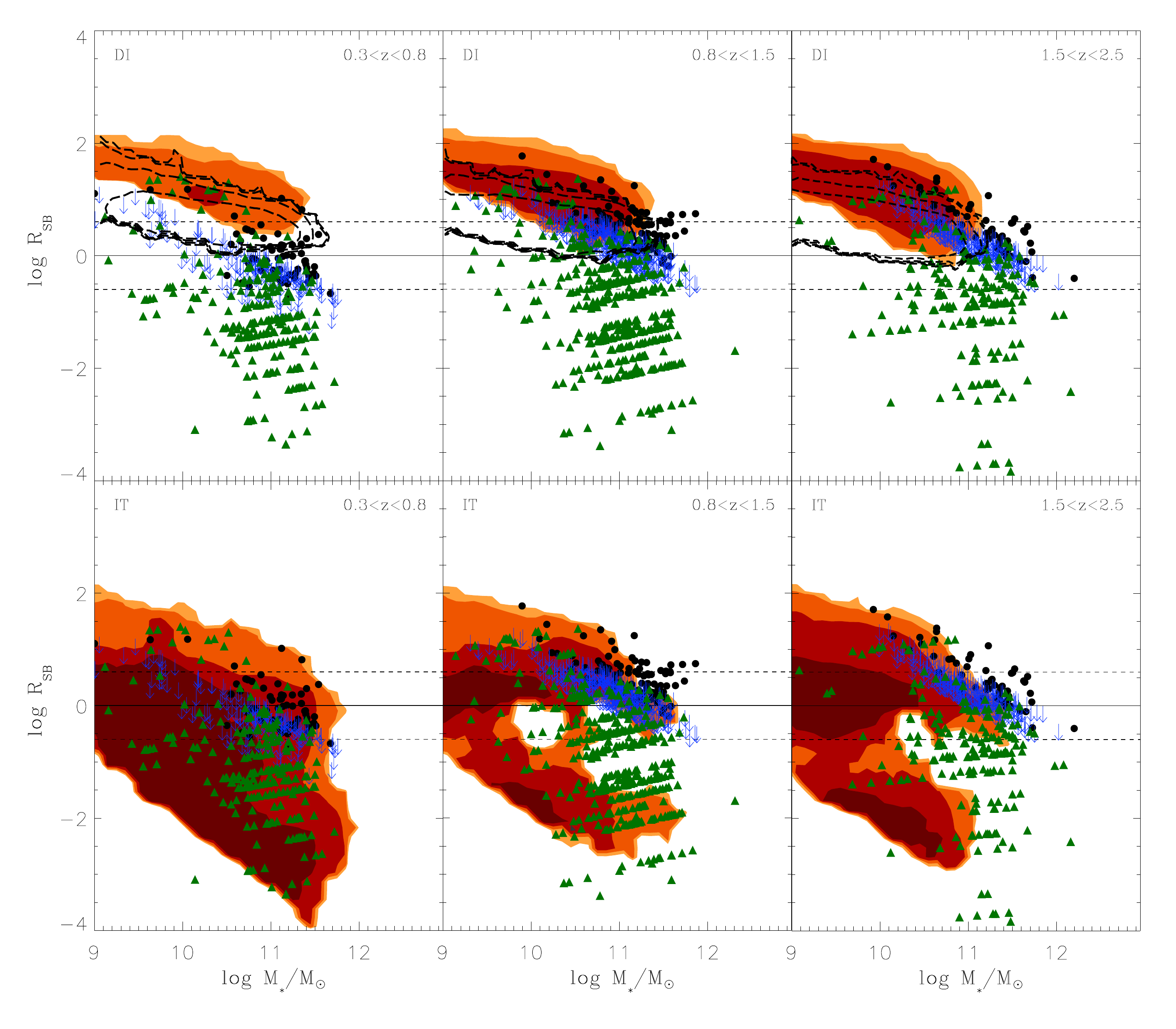}
\caption{AGN host galaxy starburstiness $R_{SB}$ as a function of $M_*$ for DI scenario (upper panels, solid contours for the fiducial model and black dashed contours for $\alpha=2$) and IT scenario (lower panels) in three different redshift bins. Only AGN with luminosity $L_X > 10^{42.2}$ erg s$^{-1}$ are considered. The filled contours correspond to equally spaced values of the density (per Mpc$^3$) of model AGN in logarithmic scale: from $10^{-7}$ for the lightest filled region to $10^{-4}$ for the darkest. The data points indicate the XMM-COSMOS AGN with $L_X > 10^{42.2}$ erg s$^{-1}$. Circles
and arrows indicate AGN with SFR derived from $L_{FIR}$, triangles indicate AGN with SFR derived from SED fitting. Solid lines show the position of the galaxy main sequence, dashed lines denote the limits of the starburst and passive regions, defined as $R_{SB} > 4$ and $R_{SB} < 1/4$, respectively.}
\label{ssfragn}
\end{figure*}

The host star formation activity can be effectively studied by using the scaling relation that connects SFR and galaxy stellar mass. Normal star-forming galaxies have recently been found to lie along a main sequence that is characterized by a typical redshift-dependent value of the specific star formation rate ($SSFR \equiv SFR/M_*$) with a small scatter ($\sim 0.3$ dex) (Brinchmann et al. 2004, Noeske et al. 2007, Elbaz et al. 2007; Daddi et al. 2007, 2009, Santini et al. 2009, Salim et al. 2007, Stark et al. 2009, González et al. 2011, Rodighiero et al. 2011,2014). 
Starburst galaxies are characterized by higher SSFR values, lying above the main sequence, while passive galaxies populate the region well below the main sequence.

Figure \ref{ssfragn} shows the starburstiness parameter ($R_{SB}$) as a function of stellar mass for galaxies hosting AGN with luminosity $L_X > 10^{42.2}$ erg s$^{-1}$ in three different redshift bins for the IT and DI scenarios. The starburstiness  parameter is defined as $R_{SB} \equiv SSFR/SSFR_{MS}$, where the subscript MS indicates the typical value for main-sequence galaxies; it measures the excess or deficiency in SSFR of a star-forming galaxy in terms of its distance from the galaxy main sequence. We plot the starburstiness instead of the $SSFR$ because of the well-known fact that hierarchical clustering models, while they succeed in reproducing the slope and the scatter of the $SSFR-M_*$ relation, underpredict its normalization (Daddi et al. 2007; Davé 2008; Fontanot et al. 2009; Damen et al. 2009; Santini et al. 2009; Weinmann et al. 2011; Lamastra et al. 2013a). Using the starburstiness, both the model predictions and observational data were normalized to their MS values, and the problem of the different normalizations of $SFR_{MS}$ is thus avoided. For the observational data, we  used the best-fit of the galaxy main sequences obtained by Santini et al. (2009) in similar redshift intervals, while for model galaxies we derived in each redshift bin the predicted galaxy main sequence following the procedure described  in Lamastra et al. (2013b).

In our previous paper (Lamastra et al. 2013b) we studied the distribution in the $R_{SB}-M_*$ diagram of galaxies hosting AGN with high luminosities ($L_X > 10^{44}$erg s$^{-1}$) for the IT model alone. 
We found that the starburst region ($R_{SB} >4$ Rodighiero et al. 2011; Sargent et al. 2012) is populated by AGN host galaxies  that are dominated by the burst component of star formation ($SFR_{b,I} > SFR_q$), while the MS ($1/4 < \,R_{SB} <  4 $) is equally populated by AGN hosts where the quiescent SF dominates ( $SFR_q > SFR_{b,I}$) or where the burst SF dominates ($SFR_{b,I} > SFR_q$). We also  found that host galaxies dominated by burst SF do not only occupy the region above the main sequence, but they reside also below, in the passive region ($R_{SB} < 1/4$). This is because the passive region is populated by massive galaxies
that are formed in biased region of the primordial density field, which have already converted most of their gas into stars as
a result of high-z interactions and merging events: because only
a small amount of gas is left in such massive galaxies, quiescent SF is strongly quenched and burst SF dominates.

Here, we extend the analysis of Lamastra et al (2013b) to the class of  less luminous AGN ( $L_X < 10^{44}$ erg s$^{-1}$) and also to AGN triggered by DIs. 
We note the DI scenario is characterized by a distribution different from that of the IT mode: AGN host galaxies mainly populate the starburst region and the upper part of the main sequence, and they are constrained to a tight region. Moreover, they completely fail to populate the passive region. This does not change even if we consider the case with lowered normalization ($\alpha=2$): AGN hosts are now shifted towards lower values of the starburstiness parameter (due to a less prominent SF activity), but they still do not populate the passive region and mainly reside in the main sequence.

The explanation of this different behaviour strictly follows the one given for the CMD. In the DI scenario, AGN are found in disky, gas-rich, late-type galaxies, and the high values of $SFR_{b,D}$ associated with AGN activity ($SFR_{b,D} = 100 \dot{M}_{BH} $) confine most of them to the upper part of the plot, in the starburst region and in the main sequence. 
The lack of AGN hosts in the lower part of the plot instead points out a relevant piece of information: even if CMD cannot always
be considered a good tracer of SF activity, the lack of AGN in the red sequence of the CMD noted above really means that DI mode is not able to trigger AGN activity in passive non-star-forming hosts.

We compared our predictions with a sample of X-ray selected AGN from the XMM-Newton survey
of the COSMOS field (Scoville et al. 2007, Cappelluti et al. 2009, Brusa et al.
2010).  The stellar masses of the XMM-COSMOS AGN were derived by Santini et al. (2012) and Bongiorno et al.
(2012)  by fitting the observed SEDs with a two-component
model based on a combination of AGN and host galaxy templates. Their SFRs are derived both from an SED-fitting technique (only for the obscured AGN subsample) and from their far-infrared emission, as described in detail in Lamastra et al. (2013b). While both IT and DI modes could be responsible for hosts observed in the main sequence or in the starburst region, DIs cannot reproduce hosts in any redshift bin in the passive region, where most AGN are found.


Given this picture, we can try to assess whether the lack of DI-driven AGN in the passive region might be alleviated by varying the values of the two main tunable quantities of the HQ11 model ($\alpha$, $SFR_{b,D}$)
or if it constitutes a robust feature of the DI scenario implemented in our SAM. We stress that the values of these two quantities \textit{are not completely freely tunable}, but can vary only within specific ranges (as explained in Sects. 2.1 and 2.2) owing to some uncertainties of the HQ11 model.

First of all, we can study the effect of assuming a different value for the $SFR_{b,D}$ associated with AGN activity in the DI scenario: a lower value for the ratio $SFR_{b,D}/\dot{M}_{BH}$ would move AGN hosts towards lower values of the starburstiness parameter. We recall that the value of burst star formation $SFR_{b,D}=100 \dot{M}_{BH}$ follows from the HQ11 model; this ratio also agrees with the observational results from Silverman et al. (2009), who have studied the correlation between $\dot{M}_*$ and $\dot{M}_{BH}$ in galaxies hosting an AGN in isolated (non-merging) systems, and with that resulting from simulations (Hopkins \& Quataert 2010, Bournaud et al. 2011). As explained in Sect. 2.2, the value $SFR_{b,D}=100 \dot{M}_{BH}$ already represents a lower limit for the burst SF within the assumptions of the HQ11 model, obtained considering the contribution to the star formation rate only from the inner region of the galaxy where the BH dominates the potential. If we were also to take the outer region into account, we would obtain higher values of the $SFR_{b,D}$, which would
move AGN hosts even more deeply into the starburst region and worsen the comparison with data. To be more quantitative, if we were to consider our fiducial model and fix the value of the ratio $SFR_{b,D}/\dot{M}_{BH}=10^3 $ (similar to the value of the local $M_{bulge}-M_{BH}$ relation), the position of AGN hosts in the $R_{SB}-M_*$ plane would be shifted upwards by 1 dex; as a consequence, all AGN hosts would reside in the starburst region, and the match with observational data would become extremely poor.

On the other hand, we could try to decrease the absolute value of the $SFR_{b,D}$ by lowering the normalization of the mass inflow predicted by the HQ11 model even more by changing the parameter $\alpha$ in Eq. \ref{hopkins}. As we recalled in Sect. 2.1, the constant $\alpha$ parametrizes several uncertainties related to some of the basic assumptions of the HQ11 model, and typical values of $\alpha$ are in the range of $\alpha = 0.1 \sim 5$. The two cases with $\alpha=10$ (the fiducial case) and $\alpha=2$ shown in Fig. 3 is expected to span all the possible reasonable values of the normalization of the inflow because a lower normalization would drastically affect the AGN luminosity function, with DIs not being able to reproduce it in almost any acceptable range of redshift and luminosity. Nevertheless, even if we were to take $\alpha=0.1$, the $SFR_{b,D}$ would be diminished only by two orders of magnitude compared to the fiducial case, clearly not enough to move DI AGN deep into the passive region. 


Finally, it is important to stress that even if we were to take $SFR_{b,D}=0$, we would not be able to move AGN hosts below the main sequence, since the quiescent component of star formation (Eq. \ref{quie}) in AGN  triggered by disk instabilities would begin to dominate the total star formation of the host galaxies. In fact, the criterion for the onset of disk instabilities (Eq. \ref{efstathi}) implies that DI-driven AGN are activated only in disky and gas-rich galaxies where the $SFR_{q}$ is high. Since galaxies dominated by the quiescent component of star formation populate the galaxy main sequence (Lamastra et al. 2013a), the SSFR of AGN hosts would not be able to reach values lower than $SSFR_{MS}$. This implies that the \textit{\textup{lack of AGN in the passive region for the DI scenario is quite insensitive to the details of the HQ11 model for the mass inflow and the assumptions made to compute the star formation associated with the AGN activity}}; it rather depends on the criterion for the onset of disk instabilities, which prevents AGN activity from being triggered in bulge-dominated, passive galaxies. 


\begin{figure*}
\centering
\includegraphics[scale=0.35]{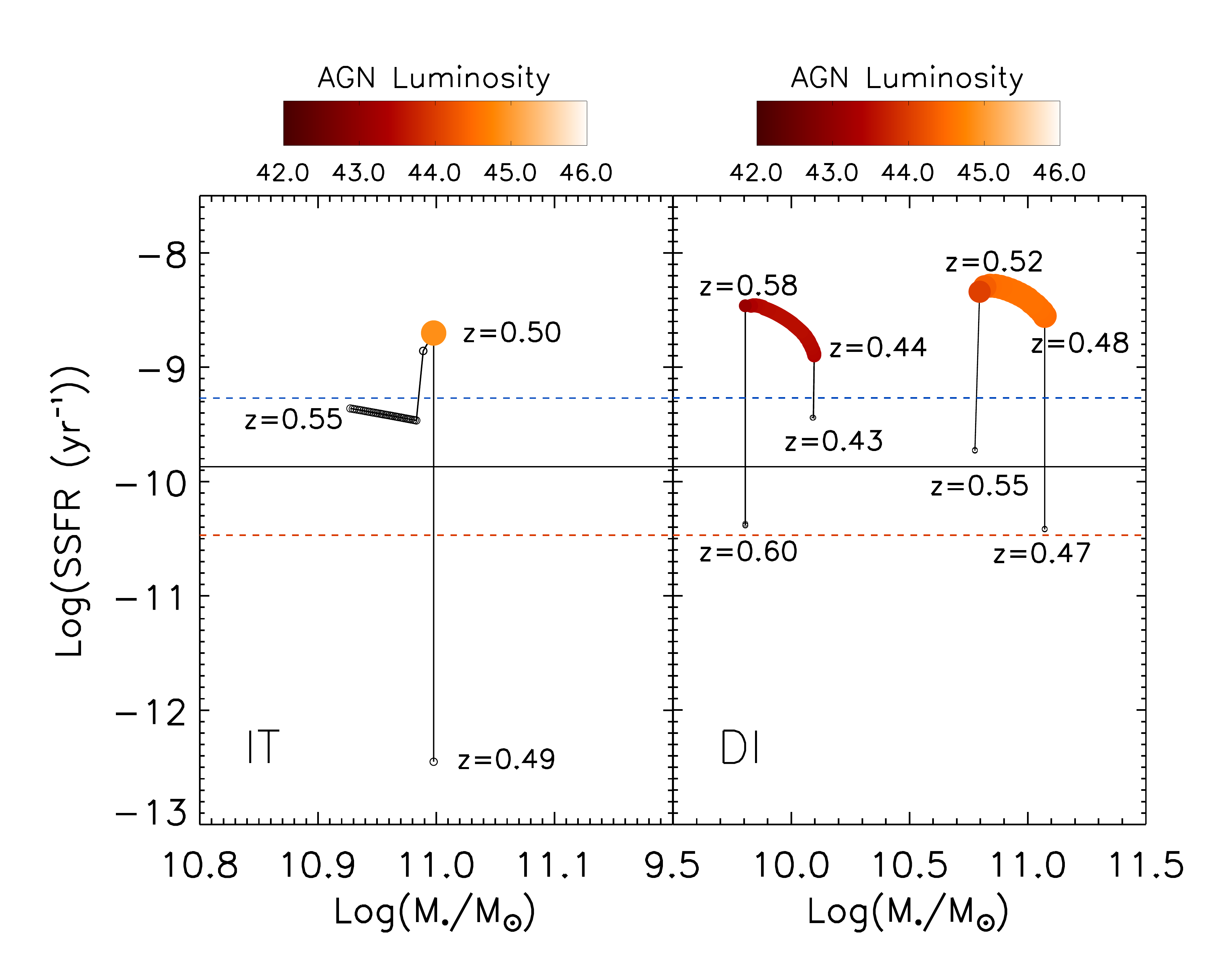}
\caption{Path followed by typical AGN hosts in the $SSFR-M_*$ plane. The left panel refers to the IT scenario, the right panels to the DI mode. The main sequence computed from the model is represented by the black line (computed in the redshift range $0.5 < z < 0.8$), while the blue and the red dotted lines mark the separation between the main sequence and the starburst/passive region. The position of AGN hosts is represented by a circle; if the circle is black, no AGN activity is present in the host, or AGN luminosity is below the threshold ($L_X > 10^{42} $ erg s$^{-1}$). AGN activity above the threshold is represented by a coloured circle (the colour code is shown above each plot, and the AGN luminosity refers to X-ray luminosity in the 2-10 keV band). We also show the corresponding redshift
near some circles; the hosts have been selected at redshift $z=0.5$. At $z=0.5$, AGN are characterized by luminosities $Log \, L_X= 44.9$ (left panel), $Log \, L_X= 44.6$ (right path, right panel) and $Log \, L_X= 43.3$ (left path, right panel).}
\label{path}
\end{figure*}

To further investigate the different distributions associated with the two modes, we also studied the typical paths followed by AGN hosts in the $SSFR-M_*$ plane. Figure \ref{path} shows a few selected paths followed by galaxies hosting AGN that at redshift $z=0.5$ are found to reside in the starburst region (defined as $SSFR > 4\,  SSFR_{MS}$). Two of the selected AGN are characterized by very similar luminosities ($Log \, L_X= 44.9$ for the IT-driven AGN, and $Log \,L_X= 44.6$ for the DI-driven AGN represented by the right path in the right panel), while the other path refers to a DI AGN host characterized by a lower AGN luminosity and an higher duty cycle. We show the paths starting from their position in the main sequence, following the evolution of AGN hosts down to low redshifts (corresponding to higher stellar mass content). For each time step, the host position in the plot is represented by a circle, while its evolution is connected by a black line. 

The paths can provide several insights about the differences between the two feeding modes. Firstly, one of the differences that can be inferred from the plots concerns the AGN feedback. In the IT scenario, AGN feedback is very efficient in removing the gas from the host, quenching star formation and preventing further AGN activity (left panel of Fig. \ref{path}). As a result, the SSFR of the host galaxy drops by several orders of magnitude, and the galaxy moves to the passive region. In case of disk instability, instead, AGN feedback is less efficient: since DIs activate only in disky and gas-rich galaxies, it is more difficult for the blast wave to expel all the gas content $M_c$  from the galaxy, and a substantial star formation activity remains even after the end of the AGN burst phase.


Secondly, another striking difference between the two feeding modes that can be inferred from Fig. \ref{path}  concerns the AGN duty cycle. In the IT scenario, the duration of AGN phase is imposed by the duration of the galaxy interaction, and typical values are in the range $\sim 10^7 - 10^8$ yr (see, e.g, Shankar et al. 2009). For the DI scenario, in contrast, AGN activity can last for a longer period of time (see left path of the right panel of Fig. \ref{path}, where AGN shines on time scales of $\sim $ Gyr), up to ten times more than IT mode. In this case, there is no specific limit for the activity duration imposed by external encounters: if the conditions for triggering AGN activity are satisfied by the host, AGN can shine for a long time. This result seems to be supported by recent simulations
as well: for instance, Bournaud et al. (2011) studied disky galaxies at $z \sim 2$ with aimed and isolated galaxy simulations and
found that DIs are able to funnel half the gas content of galaxy onto the central BH in $\sim $2 Gyr, leading to inflows similar to those obtained during major mergers, but spread over longer periods of time. 

Finally, we note that lowering the normalization of the accretion rate described by Eq. \ref{hopkins} for the
DI\ scenario causes a general lengthening of the AGN duty cycle, and with respect to the fiducial case ($\alpha=10$), there is a substantial increment of the number of AGN that can shine for a few Gyr (especially those with low luminosities). This is due to the reduced value of $SFR_{b,D}$ associated with AGN activity: since the new stars produced during the burst phase eventually contribute to the stability of the disk, reducing the value of $SFR_{b,D}$ means that it will take more time for the new stars to stabilize the disk. This leads to an overproduction of very faint AGN for the model with $\alpha=2$. This problem might be solved by including new and more detailed conditions in the SAM for the stability criterion for the onset of disk instabilities, or by better modelling the physics of the burst SF related to AGN activity. This problem goes beyond the aims of this paper, however.


\subsection{$SFR-L_{BOL}$ relation}
\begin{figure*}
    \centering
\includegraphics[scale=0.65]{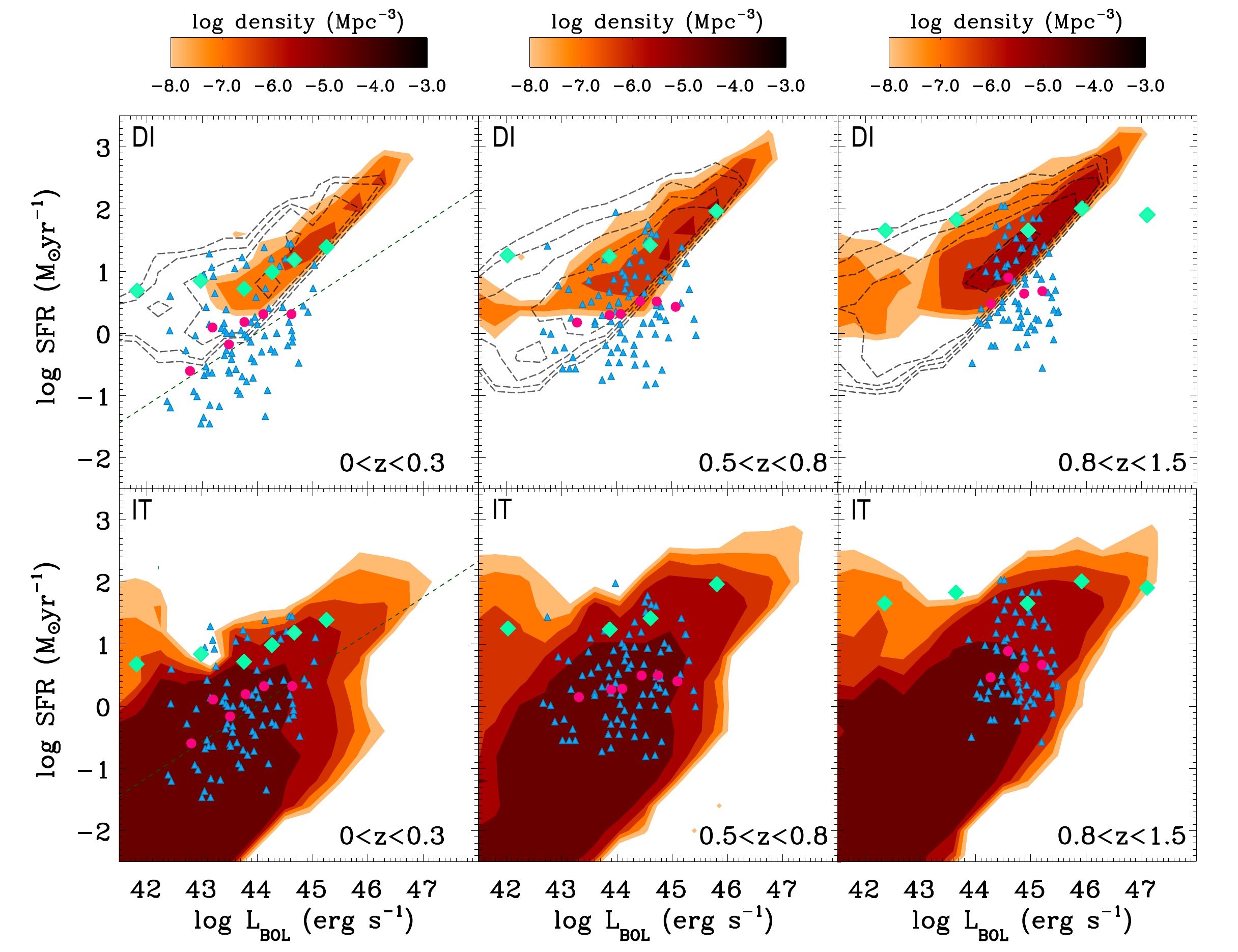}
\caption{SFR versus $L_{bol}$ for three different redshift bins. The upper panels refer to the DI scenario, the lower to the IT mode. The contour plot represents the AGN number density (per $Mpc^3$) as predicted by the model. The predictions for $\alpha=2$ for the DI scenario are represented by the dashed contour plot.  The dashed line is the relation obtained by Netzer
(2009); pale-blue diamonds are the observational results from Rosario et al. (2012). Blue triangles are data points from Azadi et al. (2014), with purple circles representing median SFR values in different luminosity bins.}
\label{sfrlbol}
\end{figure*}

The correlation between AGN luminosity and the host star formation rate could provide useful hints for the AGN triggering mechanisms. However, most observational studies that have tried to assess this correlation have revealed a complex situation, mainly due to observational bias and/or AGN variability (Hickox et al. 2014, Chen et al.  2013). In the local Universe, Netzer et al. (2009) found a strong correlation between the luminosity at  60 $\mu$m (where the far-infrared emission from cold dust heated by the UV radiation of massive young stars has its peak) and the AGN luminosity, for optically selected AGN  over more than five orders of magnitude in luminosity.  According to Rosario et al. (2012), the luminosity at  60 $\mu$m  is correlated with the AGN luminosity for  AGN with X-ray luminosities  $\rm {L_{X}}\gtrsim10^{43}$ erg s$^{-1}$ at $z<1$, while no correlation is found for AGN of the same luminosity at higher redshift and for lower luminosity AGN.  Rovilos et al. (2012) found no significant correlation in low-luminosity AGN with $L_X < 10^{43.5}$ erg s$^{-1}$ at $z < 1$ either; instead, they found a substantial correlation for higher X-ray luminosities at $z > 1$. In contrast, Mullaney et al. (2012a) found no evidence of any correlation between the X-ray and infrared luminosities of AGN with  $\rm{L_{X}}=10^{42}-10^{44}$ erg s$^{-1}$ up to $z=3$. However, the correlation arises at $z=$1-2 when stacked X-ray emission of undetected sources is taken into account (Mullaney et al. 2012b). Similarly, Azadi et al.(2014) found no evidence of a correlation between SFR and the instantaneous AGN X-ray luminosity for AGN with $\rm{L_{X}}=10^{41}-10^{44}$ erg s$^{-1}$ at $0.2< z< 1.2$, but they found a weak correlation when the mean $L_X$ of detected AGN is considered.

Figure \ref{sfrlbol} shows the $SFR-L_{bol}$ relation predicted by the  DI (upper panels) and IT scenario (lower panels) in three different redshift bins. The relation between SFR and $L_{bol}$ predicted by the IT scenario has been presented and discussed in Lamastra et al. (2013b). Because the accretion rate onto the BH is only correlated to the burst mode of star formation (Eq. \ref{burst1}) in this framework,  we found a strong correlation between  SFR and $\rm {L_{bol}}$ for luminous AGN (where the contribution of the $SFR_{b,I}$ dominates the total SFR), and a more scattered relation ($\simeq$3 orders of magnitude) for the less luminous sources owing to the larger contribution of the quiescent component of star formation (Eq. \ref{quie}) to the total SFR.

In contrast, the  DI scenario predicts a very tight $SFR-L_{bol}$ relation at each AGN luminosity. These different behaviours agree with our previous results. Because most AGN are found to reside in the starburst region and in the upper part of the galaxy MS in the DI scenario, the total star formation of the host galaxies is not dominated by the quiescent component (Lamastra et al. 2013a,b).  Although there is an additional star formation component that is unrelated to the AGN activity (Eq. \ref{burst1}) in the DI\ scenario, the strong burst of star formation related to the BH accretion (Eq. \ref{burst2}) results in an  $SFR-L_{bol}$ relation with a small intrinsic scatter. The relation shows a slightly greater scatter (but still lower than the IT scenario) if we consider the prediction for the DI scenario with lowered normalization of the mass inflow ($\alpha=2$): in this case, AGN hosts reside mainly in the SF main sequence, and the contribution of the quiescent component to the total SFR increases, broadening the relation.

Since the tightness of the $SFR-L_{bol}$ relation for the DI scenario constitutes the most striking difference between the two modes, we can try to assess the robustness of this result with respect to changes of the two parameters of the HQ11 model ($\alpha$, $SFR_{b,D}$) or its dependence on the Efstathiou criterion, similarly to what we
reported in the previous section. For instance, diminishing the ratio $SFR_{b,D}/ \dot{M}_{BH}$ or assuming lower values for the normalization of the mass inflow would be reflected in a decrement of the contribution of the $SFR_{b,D}$ to the total SFR; as a consequence, the $SFR-L_{bol}$  relation would be characterized by a larger scatter, reaching lower values of the SFR. However, since we have already noted that in the HQ11 model the ratio $SFR_{b,D}/ \dot{M}_{BH} = 100$ and the value $\alpha=2$ represent lower limits for the burst SF and for the normalization of the mass inflow, the tightness of the relation for the DI scenario seems to be a robust result, within the uncertainties of the HQ11 model. The Efstathiou criterion here has a small role in determining the shape and the tightness of the relation; since it implies that DI-driven AGN are only activated in disky and gas-rich galaxies where $SFR_{q}$ is high, its main effect on the relation is to prevent AGN hosts from populating the lower part of the plot, which is characterized by low SFR values. Ultimately, we also note that assuming a time delay between the peak of $SFR_{b,D}$ and the AGN activity would not appreciably alter the scatter of the relation for the DI scenario. This is because the typical time delays inferred from observations (Davies et al. 2007, Wild et al. 2010) and N-body simulations (Hopkins et al. 2012) are shorter (of the order of $\sim 10^7$ yr on pc scales and $\sim 10^8$ yr on kpc scales) than the typical lifetime of an AGN in the DI scenario (up to $\sim$ Gyr, see Sect. 3.3). As a consequence, the time interval where Eq. 10 would not be valid would be very short, and most AGN hosts would remain tightly constrained in the plot.


We compare our predictions with the observational data from Rosario et al. (2012) in the same redshift bins. The data points do not represent single sources but rather mean trends, obtained by
combining fluxes from detections and stacks of undetected sources in the Herschel-PACS bands for a sample of X-ray selected AGN.  We obtained SFRs from rest-frame luminosities at 60 $\mu$m,  following the procedure described in Lamastra et al. (2013b). We also compare our predictions with observational data from Azadi et al. (2014) in similar redshift bins; these authors studied a sample of low-to-intermediate luminousity non-broad-line AGN from the PRIMUS spectroscopic redshift survey (Coil et al. 2011; Cool et al. 2013). For these data, we show both the individual data points of the sample and the points representing median SFRs in different luminosity bins. X-ray luminosities were converted into bolometric luminosities using the bolometric correction by Marconi et al.(2004). Finally, we plot the observed relation for local optically selected AGN
for the lowest redshift bin (Netzer 2009).

Both IT and DI modes agree reasonably well with data points from Rosario et al. (2012), while the DI scenario
is slightly above the observational relation from Netzer et al. (2009) at low redshift; DIs are also slightly above the median points from Azadi et al., even though they are able to reproduce a substantial fraction of the observed AGN hosts. Unfortunately, at present the observational situation is still too uncertain due to observational bias, instrumental limitation in sensibility and sample selection effects, to use the $SFR-L_{bol}$ relation and its scatter to effectively distinguish between the two modes. We also note we did not include any analytical prescription accounting for AGN variability in our
SAM. This variability is considered one of the possible causes responsible for hiding the connection between AGN luminosity and star formation activity, and might affect the scatter for the two scenarios,
additionally broadening the predicted relations.

Because the relations predicted by the model are quite different for the two scenarios especially at low-intermediate AGN luminosities, we stress that the scatter of $SFR-L_{bol}$ relation could represent a crucial diagnostic, and future observations might be able to effectively pin down the dominant AGN accretion mode.

\subsection{Environmental dependence: AGN in groups and clusters}
Finally, we studied the fraction of active galaxies in groups and clusters. We expect a different environmental dependence for the two modes. For the IT mode, interactions should be enhanced in denser environment because galaxies are assumed to be closer to one another, leading to more frequent merging or fly-by events than in the field. The opposite is expected to occur for DIs. We have already stressed that they can occur only in disky, gas-rich galaxies that are characterized by young stellar populations; this is at odds with the typical galaxies found in groups or clusters, where the frequent interactions at high redshifts have led to a population of red, bulge-dominated and passive galaxies. 

\begin{figure*}
\centering
\includegraphics[scale=0.46]{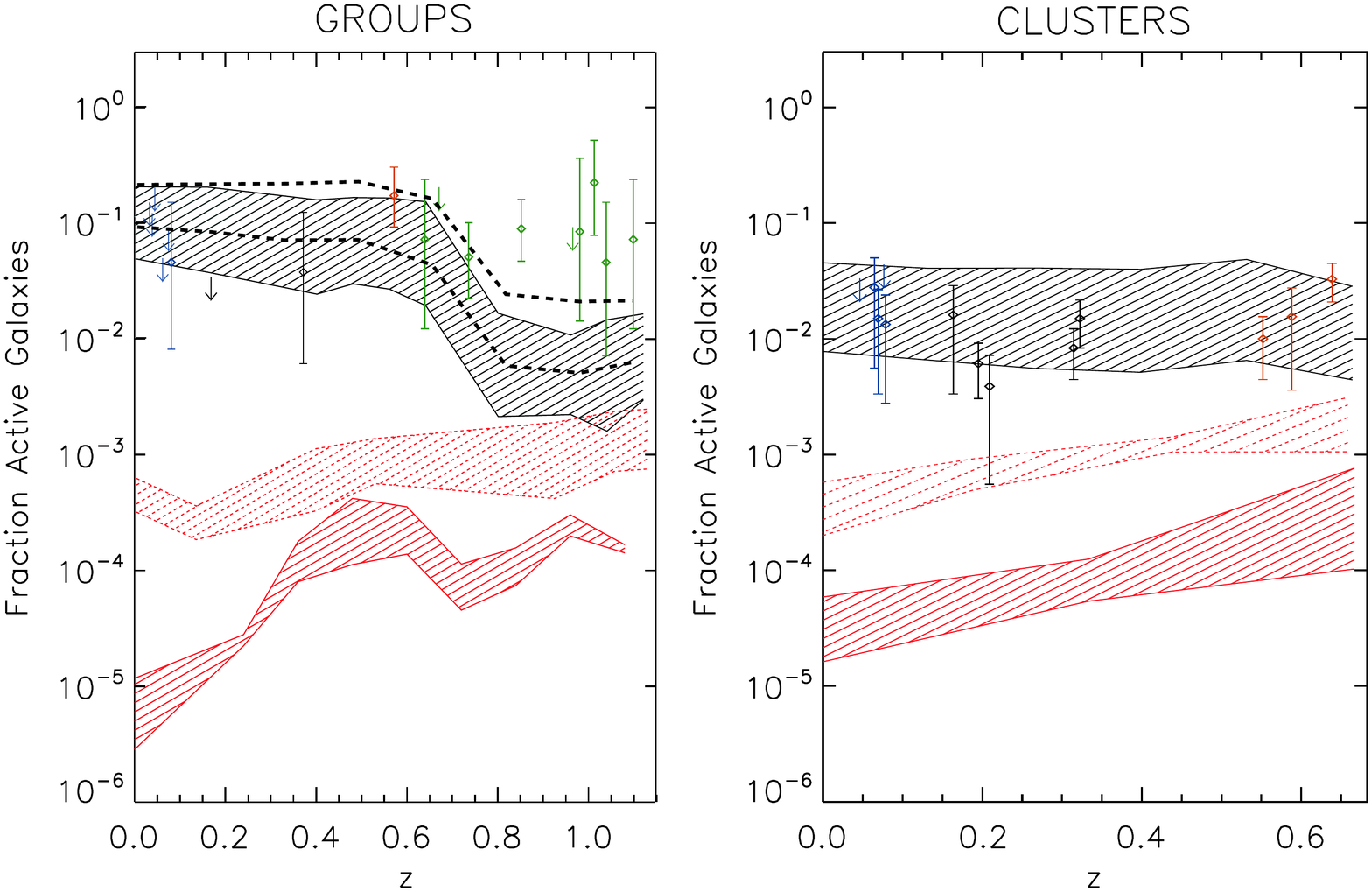}
\caption{Predicted AGN fraction in groups ($ 10^{13.5}  < M < 10^{14.3} \msun$, left panel) and clusters ($M > 10^{14.5} \msun$, right panel) for AGN $L_{X} > 10^{42}$ erg s$^{-1}$, hosted by galaxies brighter than $M_R<-20$. The predictions of our model are represented by shaded regions: black for IT and red for the fiducial case for DI. The shaded region with red-dashed lines represents the prediction for the DI scenario with decreased normalization $\alpha=2$, the two black-dashed lines in the groups plot represent the limits of the prediction for the IT scenario obtained with a slightly brighter threshold for galaxies ($M_R<-21$, see text for more details). Green symbols are data from Pentericci et al. (2013), red symbol from Eastman et al. (2007), black symbols from Martini et al. (2009), and blue symbols from Arnold et al. (2009). Uncertainties and upper limits are computed by using the low-number statistics estimator (at 1 $\sigma$) by Gehrals (1986).}
\label{cluster}
\end{figure*}
Figure \ref{cluster} shows the predicted fraction of AGN in groups (left panel) and clusters (right panel). Our predictions are compared with data from Pentericci et al. (2013); in addition
to showing their data for a number of groups and small clusters, the authors here provide a wide collection of data from the literature for groups and massive clusters. To be consistent with observational selection criteria, we considered AGN with luminosity $L_{X} > 10^{42}$ erg s$^{-1}$, hosted by galaxies brighter than $M_R < -20$. The classification into groups and clusters of the observational data we compared our results with is based on the value of dispersion velocity $\sigma$, with groups and small clusters with $350 < \sigma < 700$ km s$^{-1}$, while the most massive clusters are characterized by $\sigma > 700$  km s$^{-1}$. However, in our SAM we considered as groups structures with masses in the range $ 10^{13.5}  < M < 10^{14.3} \msun$ and as clusters structures with mass $M > 10^{14.3} \msun$. We note that selecting structures according to their mass makes little difference with data: in fact, the masses of groups or small clusters with $\sigma < 700$ km s$^{-1}$ selected by Pentericci et al. are of the order of 0.5 to few times $10^{14} \, \msun$ (see also Salimbeni et al. 2009). The predictions for the two modes are shown as shaded regions (see the figure caption for more details), up to redshift $ z \sim 1$ for groups and $z \sim 0.7$ for clusters. To only
select AGN with magnitude $M_R < -20$, we made use of the magnitude provided by our SAM, obtained by convolving galaxies SF history with spectral energy distribution from stellar population synthesis models, and of an observational magnitude-to-galaxy mass relation. This observational relation (Grazian et al., in prep.) was fitted with a straight line, and the contours of the shaded regions represent the upper and the lower limits of the fit.

As expected, the fraction of AGN predicted in the DI scenario is lower than in IT mode. Interestingly, we note that in the range of redshift where the two plots overlap ($ z\lesssim 0.6$), the fraction of AGN predicted by IT in clusters is slightly lower than in groups; as pointed out by Pentericci et al., this is due to the higher velocity dispersion in clusters, which causes interactions to be less effective (this can be noted from Eq. \ref{fdest}, where an increase of the relative velocity $V_{rel}$ diminishes the fraction of gas that is destabilized during the interaction). For the DI scenario, $\alpha=2$ predicts a greater abundance of AGN than in the fiducial case ($\alpha=10$): this is mainly due to the lengthening of the AGN duty cycle, which
is particularly effective here since we also considered AGN with low luminosity ( $Log \, L_X \sim 42$). By comparing our predictions with data, we note that only the IT scenario is able to reproduce the observed AGN fraction (although with slight scatter in groups at high redshift), while DIs constantly predict an AGN abundance lower than observed, even by several orders of magnitude.

A possibile explanation for the discrepancy in groups
at high redshift for the IT scenario might be  the known problem  of semi-analytic models that tend to overestimate the number of galaxies at the faint end of the luminosity function. This problem has been discussed in Pentericci et al. (2013) and in Salimbeni et al (2008) and it is observed at the magnitude limit used in this study ($M_R=-20$); reducing the number density of faint objects would result in an higher AGN fraction, which might alleviate the observed discrepancy. For comparison purposes, we also plot in Fig. \ref{cluster} the AGN fraction for the IT scenario obtained by selecting slightly brighter objects($M_R<-21$), this causes the AGN fraction to increase (even if it is still not enough to match the observational data).

A detailed investigation of all the observational biases affecting the measure of AGN fraction in groups or clusters goes beyond the aims of this work. Indeed, this measure might be affected by several uncertainties: for example, the uncertainties related to the estimate of masses or dispersion velocities and the relative classification into groups or clusters, the test used to assess the virialized status of structures or the diagnostic used for selecting AGN. In this respect, the discrepancy at high redshift between the IT scenario and observational data should not be considered as conclusive, and more data are needed before the predictions for the IT scenario can be considered incorrect. Here, the important point to stress is that Fig. \ref{cluster} clearly shows that DIs are unable to reproduce the observed AGN fraction in groups or clusters (for $z \lesssim 1$ and $z \lesssim 0.7$), even accounting for all the possible uncertainties related to the observational estimates. This is a natural consequence of the results obtained in the previous sections. DI mode is indeed able to trigger AGN activity only in a very specific class of objects, which must be gas-rich, medium-sized, actively star-forming disky galaxies; these are likely to be found in the field rather than in groups or clusters. In contrast, IT mode does not have all these restrictions, and thus it is the only mode able to effectively trigger AGN activity in denser environments.

\section{Conclusions}
Using an advanced semi-analytic model (SAM) for galaxy formation (M14), we have investigated the statistical effects of different AGN-triggering mechanisms on the properties of AGN host galaxies. In particular, we have studied two different  AGN feeding modes: a first one where AGN activity is triggered by disk instabilities (DI) in isolated galaxies, and a second one where AGN activity is driven by galaxy interactions (IT).
We investigated the effects of the two scenarios separately, focusing on the properties of AGN host galaxies and on the AGN  environmental dependence, to single out the regimes in which they might be responsible for triggering AGN activity. 
We obtained the following results:

\medskip

$\bullet$  both the DI model and IT model are able to account for the observed abundance of AGN host galaxies  with $M_* \lesssim 10^{11} \msun$. For more massive hosts, the DI scenario predicts a much lower space density than the IT model in every redshift bin, lying below the observational estimates for redshift $z>0.8$. This is because in hierarchical clustering scenarios 
the most massive galaxies are characterized by low gas fraction and a high B/T ratio, which strongly suppresses the BH mass inflow predicted by the DI scenario. However, even if the predictions of the two scenarios
concerning the abundance of the massive hosts are quite different, the observational estimates of the  host galaxy stellar mass function are still too affected by several uncertainties  to effectively regard this test as conclusive. 

$\bullet$ The analysis of the colour-magnitude diagram (CMD) of AGN hosts at $z < 1.5 $ can provide a good observational test to effectively distinguish between DI and IT mode: in this  redshift range the model predictions for the distributions in the CMD are considerably different for the two scenarios. While DIs are expected to yield AGN host galaxy colours skewed towards bluer colours, in the IT scenario most hosts at low redshift reside in the red sequence and gradually populate the blue cloud at higher redshifts $ 1.5 < z < 2.5$. This is because DIs only occur in disky and gas-rich galaxies, which are actively forming stars, while interactions can also occur in old red passive galaxies. Moreover, in the DI scenario the large burst of SF associated with the AGN activity further contributes to move DI hosts to the blue cloud.  This result fosters future unbiased studies of the distribution of AGN hosts in the CMD diagram.

$\bullet$ The DI scenario predicts that  the distribution of AGN hosts in the starburstiness ($SSFR/SSFR_{MS}$) - $M_*$ plane is restricted in a tight region between the upper part of the main sequence (MS) and the starburst region ($SSFR/SSFR_{MS} > 4$, Rodighiero et al. 2011, Sargent et al. 2013, Lamastra et al. 2013a), whereas the IT scenario also predicts AGN hosts in the passive region, well below the MS ($SSFR/SSFR_{MS} < 1/4$). Thus the DI scenario misses all the AGN hosted in passive galaxies, which constitute a relevant fraction of the observed AGN. This result does not depend on either the particular model assumed to describe the mass inflow or on the value of SF burst associated with AGN activity: in fact, even if we were to ignore the burst SF ($SFR_{b,D}$) predicted by the HQ11 model and considered only the quiescent SF of AGN hosts, we would not be able to populate the passive region below the main sequence. This is due to the Efsthatiou criterion for the onset of disk instabilities (Eq. \ref{efstathi}), which implies that DI-driven AGN can only be activated in disky and gas-rich galaxies where the $SFR_{q}$ is high. Since galaxies dominated by the quiescent component of star formation populate the galaxy main sequence (Lamastra et al. 2013a), any assumptions about a possible additional contribution to the total SFR induced by AGN activity will only cause AGN to move towards the starburst region, not towards the passive region. Hence, in our model the lack of AGN in the passive region for the DI scenario is quite insensitive to the particular model for the mass inflow chosen or the assumptions made to compute the star formation associated with the AGN activity; it instead depends on the criterion for the onset of disk instabilities, which prevents AGN activity from being triggered in bulge-dominated, passive galaxies.


$\bullet$ In the DI scenario AGN activity can last for a much longer period of time (even on a time scale $\sim $ Gyr) than
in the IT mode, where the galaxies interaction sets the duration of the burst phase around $\sim 10^7 - 10^8$ yr. This result agrees with those obtained in recent simulations (Bournaud et al 2011), where disk instabilities and interactions trigger similar mass inflows, but with the former spread over periods longer
by up to ten times than in IT mode. The response of host galaxies to AGN feedback is different as well: while in the IT scenario AGN feedback is very efficient in removing the gas from the host and thus quenching star formation and preventing further AGN activity, AGN feedback is less effective in the case of disk instabilities. AGN hosts are so disky and gas rich that AGN feedback is not able to efficiently eject the gas content of the disk, so that a significant amount of star formation activity remains even after the end of the AGN burst phase.

$\bullet$ The scatter of the $SFR-L_{bol}$ relation could represent a crucial diagnostics to distinguish the different AGN triggering modes. In fact, DI scenario predicts  a lower scatter (especially at low-intermediate AGN luminosities) of the relation than the IT scenario. However, the observational relation is still too uncertain to effectively be used to unequivocally determine the AGN triggering mechanism.

$\bullet$ As for the environmental dependence, disk instabilities are not able to account for the observed fraction of AGN in groups (with dispersion velocity $350 < \sigma < 700$ km s$^{-1}$) for $z \lesssim 1$ and clusters ($\sigma > 700$ km s$^{-1}$) for $ z \lesssim 0.7$. This is because in the DI scenario AGN activity is only triggered in a very specific class of objects, which must be gas-rich, medium-sized, actively star-forming disky galaxies; these are likely to be found in the field, not in groups or clusters. In contrast, AGN activity is enhanced in denser environments in the IT mode as a result of the more frequent interactions, which produce a good match with observational data.

\medskip
The results of this work constrain the regimes in which DIs might be responsible for triggering AGN activity because they are able to provide the accretion rate needed to feed  low-to-intermediate
luminosity AGN that are hosted in medium-sized, actively star forming, blue, field galaxies. Despite the present uncertainties in the observational results that still critically affect many of the observables we have compared our results with, the picture arising from our analysis strongly disfavours DIs as the main trigger mechanism for AGN activity in red, passive galaxies.

If we wish to proceed in our understanding of AGN-triggering mechanisms, future systematic studies of the properties (colour,  SFR, $M_*$) of AGN host galaxies are clearly needed. Nonetheless, a better (and more physically justified) modelling of the DI scenario might also be valuable because it would strengthen our results, leading to new insights into the differences with the IT scenario. In this respect, we believe that the main point that should be improved in our treatment of DIs is the triggering criterion. The Efstathiou criterion, although frequently used in SAMs, is only a criterion for the global stability of the disk and might not be as successful in predicting the wide range of morphologies that a perturbed disk might assume.
We recall that our treatment of DIs \textit{\textup{does need}} a triggering criterion, since the equation for the mass inflow used in this paper (Eq. 5) only holds for perturbed systems. The HQ11 model describes the mass inflow onto the SMBH using the linear perturbation theory and under the hypothesis of an initial non axisymmetric perturbation of the system. This is achieved by considering a perturbative potential $\Phi_a$= a $\Phi_0$, where $\Phi_0$ is the axisymmetric unperturbed potential of the system. We adopted the value of a=0.3 for the perturbation amplitude (see Menci et al. 2014 for all the explicit calculations), which constitutes an upper limit to values measured in simulations (which range from $10^{-2}$ to $3 10^{-1}$, Hopkins \& Quataert 2010). This implies that Eq. 5 may only be applied to \textit{\textup{significantly perturbed}} systems, and this is why we have to include the Efstathiou stability criterion, which ensures the galaxy disk to be perturbed. Given this picture, a possible improvement of the model could consist of including no triggering condition, but instead using only Eq. 5 after releasing the hypothesis of a=0.3 and considering the full range of possible perturbation amplitudes (and possibly also different kind of perturbative potentials). This could represent an extremely intriguing improvement of the model, but  it can
only be done by considering a detailed dynamical theory that
can provide the correct perturbative amplitude (and potential) for any galaxy of the SAM. We defer this complex task to future works.

\section*{Acknowledgements} 
The authors thank Paola Santini for computing the FIR-based SFRs and Andrea Grazian for providing us the magnitude-stellar mass relation used in Sect. 3.5.
\bibliographystyle{agsm}

\begin{thebibliography}{}



\bibitem{}Aird, J., et al. 2012, ApJ, 746, 90
\bibitem{}Arnold, T. J., Martini, P., Mulchaey, J. S., Berti, A., \& Jeltema, T. E. 2009, ApJ,
707, 1691
\bibitem{}Azadi, M., Aird, J., Coil, A., Moustakas, J., Mendez, A., Blanton, M., Cool, R., Eisenstein, D., Wong, K., Zhu G. 2014 arXiv 1407.1975A
\bibitem{}Baldry, I.K., Glazebrook, K., Brinkmann, J., Zeljko, I., Lupton, 
R.H., Nichol, R.C., Szalay, A.S. 2004, ApJ, 600, 681 
\bibitem{}Bahcall, J. N., Kirhakos, S., Saxe, D. H., \& Schneider, D. P. 1997, ApJ, 479, 642
\bibitem{}Barnes, J.E., Hernquist, L.E. 1991, ApJ, 370, L65
\bibitem{}Barnes, J. E., Hernquist, L. 1996, ApJ, 471, 115
\bibitem{}Blanton, M. R., \& Roweis, S. 2007, AJ, 133, 734
\bibitem{}Bond, J.R., Cole, S., Efstathiou, G., \& Kaiser, N.,1991, ApJ, 379, 440
\bibitem{}Bongiorno, A., Zamorani, G., Gavignaud, I., et al. 2007, A\&A, 472, 443
\bibitem{}Bongiorno, A., et al. 2010, A\&A, 510, A56
\bibitem{}Bongiorno, A., Merloni, A., Brusa, M., et al. 2012, MNRAS 427, 3103
\bibitem{}Bournaud, F. et al., 2011, ApJ, 741, L33
\bibitem{}Bower R. G., Benson A. J., Malbon R., Helly J. C., Frenk C. S., Baugh C. M., Cole S., Lacey
C. G., 2006, MNRAS, 370, 645
\bibitem{}Brinchmann, J., Charlot, S.,White, S. D. M., et al. 2004, MNRAS, 351, 1151
\bibitem{}Brusa, M., Civano, F., Comastri, A., et al. 2010, ApJ, 716, 348
\bibitem{}Bruzual, G., Charlot, S. 2003, MNRAS, 344, 1000
\bibitem{}Cappelluti, N., Brusa, M., Hasinger, G., et al. 2009, A\&A, 497, 635
\bibitem{}Cardamone, C. N., Urry, C. M., Schawinski, K., Treister, E., Brammer, G., Gawiser, E., 2010a, ApJ,712 L38
\bibitem{}Cavaliere, A.,  \& Vittorini, V.  2000, ApJ, 543, 599
\bibitem{}Cavaliere, A., Lapi, A., Menci, N., 2002, ApJ, 581, L1
\bibitem{}Chen, C.-T. J., et al. 2013, ApJ, 773, 3
\bibitem{}Cisternas M., Jahnke K., Inskip K. J., Inskip 2010, in IAU Symposium Vol. 267 of IAU
Symposium, Quasars Do Not Live in Merging Systems: No Enhanced Merger Rate at z < 0.8.
pp 326
\bibitem{}Coil, A. L., et al. 2009, ApJ, 701, 1484
\bibitem{}Coil, A. L., et al. 2011, ApJ, 741, 8
\bibitem{}Cool, R. J., et al. 2013, ApJ, 767, 118
\bibitem{}Cowie, L. L., Songaila, A., Hu, E. M., \& Cohen, J. G. 1996, AJ, 112, 839
\bibitem{}Cox, T. J., Dutta, S. N., Di Matteo, T., Hernquist, L., Hopkins, P. F., Robertson, B., Springel, V. 2006, ApJ, 650, 791
\bibitem{}Cox, T. J., Jonsson, P., Somerville, R. S., Primack, J. R., \& Dekel, A. 2008,
MNRAS, 384, 386
\bibitem{}Croton, D.J, Springel, V., White, S.D.M., De Lucia, G.,Frenk, C.S., Gao, L., Jenkins, A.,
Kauffmann, G., Navarro, J.F., Yoshida, N., 2006, MNRAS, 365, 11
\bibitem{}Daddi, E., Dickinson, M., Morrison, G., et al. 2007, ApJ, 670, 156
\bibitem{}Daddi, E., Dannerbauer, H., Stern, D., et al. 2009, ApJ, 694, 1517
\bibitem{}Dale, D. A., \& Helou, G. 2002, ApJ, 576, 159
\bibitem{}Damen, M., Labbé, I., Franx, M., et al. 2009, ApJ, 690, 937
\bibitem{}Davé, R. 2008, MNRAS, 385, 147
\bibitem{}Davies, R. I., Sánchez, F. M., Genzel, R., Tacconi, L. J., Hicks,
E. K. S., Friedrich, S., \& Sternberg, A. 2007, ApJ, 671, 1388
\bibitem{}Di Matteo, T., Springel, V., \& Hernquist, L. 2005, Nature, 433, 604
\bibitem{}Disney, M.J., et al. 1995, Nature, 376, 150
\bibitem{}Eastman, J., Martini, P., Sivakoff, G., et al. 2007, ApJ, 664, L9
\bibitem{}Efstathiou G., Lake G., Negroponte J., 1982, MNRAS, 199, 1069
\bibitem{}Elbaz, D., Daddi, E., Le Borgne, D., et al. 2007, A\&A, 468, 33
\bibitem{}Fanidakis, N. et al. 2012, MNRAS, 419, 2797
\bibitem{}Ferrarese, L., \& Merritt, D. 2000, ApJ, 539, L9
\bibitem{}Fiore F. et al. 2003, A\&A,409, 79
\bibitem{}Fiore, F., et al. 2012, A\&A, 537, 22
\bibitem{}Fontanot, F., De Lucia, G., Monaco, P., Somerville, R. S., \& Santini, P. 2009,
MNRAS, 397, 1776
\bibitem{}Gebhardt, K., Bender, R., Bower, G., et al. 2000, ApJ, 539, L13
\bibitem{}Gehrels, N. 1986, ApJ, 303, 336
\bibitem{}Georgakakis A., et al., 2009, MNRAS, 397, 623
\bibitem{}Georgakakis, A. \& Nandra, K. 2011, MNRAS, 414, 992
\bibitem{}González, V., Labbé, I., Bouwens, R. J., et al. 2011, ApJ, 735, L34
\bibitem{}Grogin, N. A., et al. 2005, ApJ 627, L97
\bibitem{}H\"aring, N., Rix, H.-W. 2004, ApJ, 604, L89
\bibitem{}Hasinger G., Cappelluti N., et al. B., 2007, ApJs, 172, 29
\bibitem{}Heckman, T.M. \& Best, P.N. 2014. ARAA in press (astro-ph 1403.4620)
\bibitem{}Henriques, B.M.B., \& Thomas, P.A. 2010, MNRAS, 403, 768
\bibitem{}Hernquist, L. 1989, Nature, 340, 687
\bibitem{}Hickox, R.C. et al. 2009, ApJ, 696, 891
\bibitem{}Hickox, R. C., et al. 2014, ApJ, 782, 9
\bibitem{}Hirschmann, M., Somerville, R.S., Naab, T., Burkert, A. 2012, MNRAS 426, 237
\bibitem{}Hopkins, P. F., Richards, G. T., Hernquist, L. 2007b, ApJ, 654, 731
\bibitem{}Hopkins, P.H. et al. 2009a, MNRAS, 397, 802
\bibitem{}Hopkins, P.F., Quataert, E. 2010, MNRAS, 407, 1529
\bibitem{}Hopkins, P.F., Quataert, E. 2011, MNRAS, 411, 1027 (HQ11)
\bibitem{}Hutchings, J.B. 1987, ApJ, 320, 122
\bibitem{}Kennicutt Jr. R. C., 1998, ApJ, 498, 541
\bibitem{}Kirhakos, S., Bahcall, J.N., Schneider, D.P., Kristian, J. 1999, ApJ, 520, 67
\bibitem{}Kormendy, J., \& Richstone, D. 1995, ARA\&A, 33, 581
\bibitem{}Kocevski D. D., Faber S. M., Mozena M., Koekemoer A. M., Nandra K., Rangel C., Laird
E. S., Brusa M., Wuyts S., Trump J. R., Koo D. C., Somerville R. S., Bell E. F., Lotz J. M.,
Alexander D. M., Bournaud F., Conselice C. J., Dahlen T., 2012, ApJ, 744, 148
\bibitem{}Kormendy J., Ho L. C., 2013, ARA\&A, 51, 511 2.5
\bibitem{}Lacey, C., \& Cole, S., 1993, MNRAS, 262, 627 
\bibitem{}La Franca, F. et al. 2005, ApJ, 635, 864
\bibitem{}Lamastra, A., Menci, N., Fiore, F. and Santini, P. 2013a, A\&A 552A, 44
\bibitem{}Lamastra, A., Menci, N., Fiore, F., Santini, P. Bongiorno, A. et al. 2013b, A\&A,
559A, 56
\bibitem{}Lapi, A., Cavaliere, A., \&  Menci, N. 2005, ApJ, 619, 60
\bibitem{}Li Y., Hernquist L., Robertson B., Cox T. J., Hopkins P. F.,
Springel V., Gao L., Di Matteo T., Zentner A. R., Jenkins
A., Yoshida N., 2007, ApJ, 665, 187
\bibitem{}Lilly S. J. et al., 2009, ApJS, 184, 218
\bibitem{}Lutz, D., Mainieri, V., Rafferty, D., et al. 2010, ApJ, 712, 1287
\bibitem{}Magorrian, J., Tremaine, S., Richstone, D., et al. 1998, AJ, 115, 2285
\bibitem{}Makino J., Hut P., 1997, ApJ, 481, 83
\bibitem{}Marconi, A., \& Hunt, L. K. 2003, ApJ, 589, L21
\bibitem{}Marconi, A., Risaliti, G., Gilli, R., Hunt, L.K.,Maiolino, R., Salvati, M. 2004, MNRAS, 351, 169
\bibitem{}Martini, P., Sivakoff, G. R., \& Mulchaey, J. S. 2009, ApJ, 701, 66
\bibitem{}McConnell N. J., Ma C.-P., 2013, ApJ, 764, 184
\bibitem{}Menci, N., Cavaliere, A., Fontana, A., Giallongo, E., \& Poli, F. 2004,ApJ, 604, 12
\bibitem{}Menci, N., Fontana, A., Giallongo, E., Grazian, A.,Salimbeni, S. 2006, ApJ,  647, 753
\bibitem{}Menci, N., Fiore, F., Puccetti, S., Cavaliere, A. 2008, ApJ, 686, 219
\bibitem{}Menci, N., Gatti, M., Fiore, F. Lamastra, A., 2014 arXiv 1406, 7740 M (M14)
\bibitem{}Mihos, J. C., \& Hernquist, L. 1994, ApJ, 431, L9
\bibitem{}Mihos, J. C., \& Hernquist, L. 1996, ApJ, 464, 641
\bibitem{}Mo, H.J, Mao S., \& White, S.D.M., 1998, MNRAS, 295, 319
\bibitem{}Mullaney, J. R., et al. 2012a, MNRAS, 419, 95
\bibitem{}Mullaney, J. R., et al. 2012b, ApJL, 753, L30
\bibitem{}Nandra K. et al., 2007, ApJ, 660, L11
\bibitem{}Netzer H., \& Trakhtenbrot B. 2007, ApJ, 654, 754
\bibitem{}Netzer, H. 2009b, ApJ, 695, 793
\bibitem{}Noeske, K. G., Weiner, B. J., Faber, S. M., et al. 2007, ApJ, 660, L43
\bibitem{}Pentericci L., Castellano M., Menci N., et al., 2013, A\&A, 552,
A111
\bibitem{}Pierce C. M.et al. 2007, ApJ, 660, L19
\bibitem{}Puccetti, S., et al. 2006, A\&A, 457, 501
\bibitem{}Robertson, B., Cox, T. J., Hernquist, L., Franx, M., Hopkins, P. F., Martini, P., Springel,
V. 2006a, ApJ, 641, 21
\bibitem{}Robertson, B., Hernquist, L., Cox, T. J., Di Matteo, T., Hopkins, P. F., Martini, P., Springel,
V. 2006b, ApJ, 641, 90
\bibitem{}Rodighiero G., Daddi E., Baronchelli I., et al. 2011, ApJL,
739, L40
\bibitem{}Rodighiero, G., Renzini, A., Daddi, E., et al. 2014, ArXiv:1406.1189
\bibitem{}Rosario D. J. et al., 2012, A\&A, 545, A45
\bibitem{}Rosario, D.J. et al. 2013a, ApJ 763, 59
\bibitem{}Rosario, D.J. et al 2013b, A\&A, 560, A72
\bibitem{}Salim, S., Rich, R. M., Charlot, S., et al. 2007, ApJS, 173, 267
\bibitem{}Salimbeni, S., Castellano, M., Pentericci, L., et al. 2009, A\&A, 501, 865
\bibitem{}Salucci P., Szuszkiewicz E., Monaco P., Danese L., 1999, MNRAS, 307, 637
\bibitem{}Sanders, D. B., \& Mirabel, I.F. 1996, ARA\&A, 34, 749
\bibitem{}Santini, P., Fontana, A., Grazian, A., et al. 2009, A\&A, 504, 751
\bibitem{}Santini, P., Fontana, A., Grazian, A., et al. 2012, A\&A, 538, A33
\bibitem{}Sargent, M. T., Béthermin, M., Daddi, E., \& Elbaz, D. 2012, ApJ, 747, L31
\bibitem{}Saslaw, W.C., 1985, {\it  Gravitational Physics of Stellar and Galactic Systems} (Cambridge: Cambridge Univ. Press) 
\bibitem{}Schmidt, M. 1968, ApJ, 151, 393
\bibitem{}Scoville, N., Aussel, H., Brusa, M., et al. 2007, ApJS, 172, 1
\bibitem{}Shao, L., Lutz, D., Nordon, R., et al. 2010, A\&A, 518, L26
\bibitem{}Shankar F., Bernardi, M. \& Haiman, Z. 2009, ApJ, 694, 867 
\bibitem{}Shankar F., Weinberg D. H., \& Miralda-Escude J. 2013, MNRAS, 428, 421
\bibitem{}Shemmer O. et al. 2004, ApJ, 644, 86
\bibitem{}Silverman, J.D. et al., 2009, ApJ, 696,396
\bibitem{}Silverman, J.D. et al. 2011, ApJ, 743, 2
\bibitem{}Sijacki, D., Vogelsberger, M., Genel, S., et al. 2014, MNRAS,
submitted (arXiv:1408.6842)
\bibitem{}Soltan, A. 1982, MNRAS, 200, 115
\bibitem{}Springel, V. 2005, Nature, 435, 629
\bibitem{}Stark, D. P., Ellis, R. S., Bunker, A., et al. 2009, ApJ, 697, 1493
\bibitem{}Trakhtenbrot, B. et al. 2011, ApJ, 730, 7
610, L85
\bibitem{}Treister, E., Schawinski, K., Urry, C.M., and Simmons, B.D. 2012, ApJ, 758, L39
\bibitem{}Ueda Y., Akiyama M., Ohta K., Miyaji T., 2003, ApJ, 598,
886
\bibitem{}Ueda, Y., et al. 2008, ApJS, 179, 124
\bibitem{}Ueda, Y., Akiyama, M., Hasinger, G., Miyaji, T., \& Watson, M. G. 2014, ApJ, 786, 104
\bibitem{}Villar-Martn, M. et al. 2010, MNRAS, 416, 262
\bibitem{}Villar-Martn, M. et al. 2012, MNRAS, 423, 80
\bibitem{}Villforth, C., et al. 2014, MNRAS, 439, 3342
\bibitem{}Weinmann, S. M., Neistein, E., \& Dekel, A. 2011, MNRAS, 417, 2737
\bibitem{}Wild, V., Heckman, T., \& Charlot, S. 2010, MNRAS, 405, 933
\bibitem{}Willott C. J. et al., 2010b, AJ, 140, 546
\bibitem{}Xue, Y. Q., et al. 2010, ApJ, 720, 368
\bibitem{}Yates, M.G., Miller, L., \& Peacock, J.A. 1989, MNRAS, 240, 129
\bibitem{}Yu, Q., \& Tremaine, S., 2002, MNRAS, 335, 965

\end{thebibliography}

\end{document}